\documentclass[10pt,tightenlines,prl,showpacs,a4paper,aps,twocolumn,nofootinbib,superscriptaddress]{revtex4-2} 

\usepackage{mathtools}
\usepackage{keycommand} 
\usepackage{amsthm,amsmath,amssymb,comment,dsfont} 
\usepackage{xfrac}
\usepackage{bbold}
\usepackage[none]{hyphenat}
\usepackage{fancyhdr}
\newtheorem{theorem}{Theorem}

\usepackage{subfigure}

\newcommand{\ket}[1]{\left| #1 \right\rangle}
\newcommand{\bra}[1]{\left\langle #1 \right|}

\newcommand{\beq}{\begin{equation}}
\newcommand{\eeq}{\end{equation}}
\newcommand{\bea}{\begin{align}}
\newcommand{\eea}{\end{align}}

\usepackage[dvipsnames]{xcolor}

\definecolor{blue}{rgb}{0,0,1}
\definecolor{grey}{rgb}{0.6,0.6,0.6}

\definecolor{myurlcolor}{rgb}{0,0,0.7}
\definecolor{myrefcolor}{rgb}{0.8,0,0}
\definecolor{purple}{RGB}{128,0,128}
\definecolor{ultramarine}{RGB}{63, 0, 255}
\definecolor{medblue}{RGB}{0, 0, 100}
\definecolor{googleblue}{RGB}{34, 0, 204}
\definecolor{panblue}{RGB}{0,24,150}
\definecolor{carmine}{RGB}{150, 0, 24}
\definecolor{gray}{RGB}{150, 150, 150}

\usepackage[unicode=true,pdfusetitle, bookmarks=false,bookmarksnumbered=false,
bookmarksopen=false, breaklinks=false,pdfborder={0 0 0},backref=false,
colorlinks=true,
linkcolor=carmine,
citecolor=googleblue,
urlcolor=panblue,
anchorcolor=OliveGreen,
]{hyperref}

\usepackage[normalem]{ulem}
\usepackage{rotating}
\usepackage{floatpag}
\rotfloatpagestyle{empty}

\usepackage{xspace}
\usepackage{enumerate}
\graphicspath{{.}{./figures/}}
\usepackage{stmaryrd}
\usepackage{docmute}

\usepackage{tikzfig} 

\newkeycommand{\pointmap}[style=point][1]{\,\tikz{\node[style=\commandkey{style}] (x) at (0,-0.3) {$#1$};\node [style=none] (3) at (0, 1.0) {};\node [style=none] (3) at (0, -1.0) {};\draw (x)--(0,0.7);}\,}

\newkeycommand{\typedkpoint}[style=kpoint,edgestyle=][2]{%
\,\begin{tikzpicture}
  \begin{pgfonlayer}{nodelayer}
    \node [style=none] (0) at (0, 1) {};
    \node [style=\commandkey{style}] (1) at (0, -0.75) {$#2$};
    \node [style=right label] (2) at (0.25, 0.5) {$#1$};
  \end{pgfonlayer}
  \begin{pgfonlayer}{edgelayer}
    \draw [\commandkey{edgestyle}] (1) to (0.center);
  \end{pgfonlayer}
\end{tikzpicture}\,}

\newkeycommand{\typedkpointdag}[style=kpoint adjoint,edgestyle=][2]{%
\,\begin{tikzpicture}
  \begin{pgfonlayer}{nodelayer}
    \node [style=none] (0) at (0, -1) {};
    \node [style=\commandkey{style}] (1) at (0, 0.75) {$#2$};
    \node [style=right label] (2) at (0.25, -0.5) {$#1$};
  \end{pgfonlayer}
  \begin{pgfonlayer}{edgelayer}
    \draw [\commandkey{edgestyle}] (0.center) to (1);
  \end{pgfonlayer}
\end{tikzpicture}\,}

\newcommand{\bistateadj}[1]{\,\begin{tikzpicture}[yshift=-3mm]
  \begin{pgfonlayer}{nodelayer}
    \node [style=kpointadj, minimum width=1 cm, inner sep=2pt] (0) at (0, 0.5) {$\psi$};
    \node [style=none] (1) at (-0.75, 0.25) {};
    \node [style=none] (2) at (0.75, 0.25) {};
    \node [style=none] (3) at (-0.75, -0.5) {};
    \node [style=none] (4) at (0.75, -0.5) {};
  \end{pgfonlayer}
  \begin{pgfonlayer}{edgelayer}
    \draw (1.center) to (3.center);
    \draw (2.center) to (4.center);
  \end{pgfonlayer}
\end{tikzpicture}\,}

\newkeycommand{\pointketbra}[style1=copoint,style2=point][2]{\,%
\begin{tikzpicture}
  \begin{pgfonlayer}{nodelayer}
    \node [style=none] (0) at (0, -1.5) {};
    \node [style=\commandkey{style1}] (1) at (0, -0.75) {$#1$};
    \node [style=\commandkey{style2}] (2) at (0, 0.75) {$#2$};
    \node [style=none] (3) at (0, 1.5) {};
  \end{pgfonlayer}
  \begin{pgfonlayer}{edgelayer}
    \draw (0.center) to (1);
    \draw (2) to (3.center);
  \end{pgfonlayer}
\end{tikzpicture}\,}

\newkeycommand{\twocopointketbra}[style1=copoint,style2=copoint,style3=point][3]{\,%
\begin{tikzpicture}
  \begin{pgfonlayer}{nodelayer}
    \node [style=none] (0) at (-0.75, -1.5) {};
    \node [style=none] (1) at (0.75, -1.5) {};
    \node [style=\commandkey{style1}] (2) at (-0.75, -0.75) {$#1$};
    \node [style=\commandkey{style2}] (3) at (0.75, -0.75) {$#2$};
    \node [style=\commandkey{style3}] (4) at (0, 0.75) {$#3$};
    \node [style=none] (5) at (0, 1.5) {};
  \end{pgfonlayer}
  \begin{pgfonlayer}{edgelayer}
    \draw (0.center) to (2);
    \draw (1.center) to (3);
    \draw (4) to (5.center);
  \end{pgfonlayer}
\end{tikzpicture}\,}

\newkeycommand{\twopointketbra}[style1=copoint,style2=point,style3=point][3]{\,%
\begin{tikzpicture}
  \begin{pgfonlayer}{nodelayer}
    \node [style=none] (0) at (0, -1.5) {};
    \node [style=none] (1) at (-0.75, 1.5) {};
    \node [style=\commandkey{style1}] (2) at (0, -0.75) {$#1$};
    \node [style=\commandkey{style2}] (3) at (-0.75, 0.75) {$#2$};
    \node [style=\commandkey{style3}] (4) at (0.75, 0.75) {$#3$};
    \node [style=none] (5) at (0.75, 1.5) {};
  \end{pgfonlayer}
  \begin{pgfonlayer}{edgelayer}
    \draw (0.center) to (2);
    \draw (1.center) to (3);
    \draw (4) to (5.center);
  \end{pgfonlayer}
\end{tikzpicture}\,}

\newkeycommand{\pointbraket}[style1=point,style2=copoint][2]{\,%
\begin{tikzpicture}
  \begin{pgfonlayer}{nodelayer}
    \node [style=\commandkey{style1}] (0) at (0, -0.5) {$#1$};
    \node [style=\commandkey{style2}] (1) at (0, 0.5) {$#2$};
  \end{pgfonlayer}
  \begin{pgfonlayer}{edgelayer}
    \draw (0) to (1);
  \end{pgfonlayer}
\end{tikzpicture}\,}

\newkeycommand{\kpointbraket}[style1=kpoint,style2=kpoint adjoint][2]{\,%
\begin{tikzpicture}
  \begin{pgfonlayer}{nodelayer}
    \node [style=\commandkey{style1}] (0) at (0, -0.75) {$#1$};
    \node [style=\commandkey{style2}] (1) at (0, 0.75) {$#2$};
  \end{pgfonlayer}
  \begin{pgfonlayer}{edgelayer}
    \draw (0) to (1);
  \end{pgfonlayer}
\end{tikzpicture}\,}

\newkeycommand{\kpointmap}[style1=kpoint,style2=map][2]{\,%
\begin{tikzpicture}
  \begin{pgfonlayer}{nodelayer}
    \node [style=\commandkey{style1}] (0) at (0, -1.1) {$#1$};
    \node [style=\commandkey{style2}] (1) at (0, 0.2) {$#2$};
    \node [style=none] (2) at (0, 1.5) {};
  \end{pgfonlayer}
  \begin{pgfonlayer}{edgelayer}
    \draw (0) to (1);
    \draw (1) to (2.center);
  \end{pgfonlayer}
\end{tikzpicture}%
\,}

\newkeycommand{\sandwichtwo}[style1=point,style2=map,style3=map,style4=copoint][4]{\,%
\begin{tikzpicture}
  \begin{pgfonlayer}{nodelayer}
    \node [style=\commandkey{style2}] (0) at (0, -0.75) {$#2$};
    \node [style=\commandkey{style3}] (1) at (0, 0.75) {$#3$};
    \node [style=\commandkey{style1}] (2) at (0, -2) {$#1$};
    \node [style=\commandkey{style4}] (3) at (0, 2) {$#4$};
  \end{pgfonlayer}
  \begin{pgfonlayer}{edgelayer}
    \draw (2) to (0);
    \draw (0) to (1);
    \draw (1) to (3);
  \end{pgfonlayer}
\end{tikzpicture}\,}

\newkeycommand{\longbraket}[style1=point,style2=copoint][2]{\,%
\begin{tikzpicture}
  \begin{pgfonlayer}{nodelayer}
    \node [style=\commandkey{style1}] (0) at (0, -2) {$#1$};
    \node [style=\commandkey{style2}] (1) at (0, 2) {$#2$};
  \end{pgfonlayer}
  \begin{pgfonlayer}{edgelayer}
    \draw (0) to (1);
  \end{pgfonlayer}
\end{tikzpicture}\,}

\newkeycommand{\onb}[style=point]{\ensuremath{\left\{\tikz{\node[style=\commandkey{style}] (x) at (0,-0.3) {$j$};\draw (x)--(0,0.7);}\right\}}\xspace}

\newkeycommand{\copointmap}[style=copoint][1]{\,\tikz{\node[style=\commandkey{style}] (x) at (0,0.3) {$#1$};\draw (x)--(0,-0.7);}\,}

\tikzstyle{cdiag}=[matrix of math nodes, row sep=3em, column sep=3em, text height=1.5ex, text depth=0.25ex,inner sep=0.5em]
\tikzstyle{arrow above}=[transform canvas={yshift=0.5ex}]
\tikzstyle{arrow below}=[transform canvas={yshift=-0.5ex}]

\usetikzlibrary{shapes.multipart}

\tikzstyle{bwSpider}=[
       rectangle split,
       rectangle split parts=2,
       rectangle split part fill={black,white},
 minimum size=3.6 mm, inner sep=-2mm, draw=black,scale=0.5,rounded corners=0.8 mm
       ]
 \tikzstyle{wbSpider}=[
       rectangle split,
       rectangle split parts=2,
       rectangle split part fill={white,black},
 minimum size=3.6 mm, inner sep=-2mm, draw=black,scale=0.5,rounded corners=0.8 mm
       ]
\tikzstyle{qWire}=[line width = 1pt, color=quantumviolet]
\tikzstyle{hWire}=[line width = 1.5pt, color=higherorderred]
\tikzstyle{sWire}=[line width = 2pt, color=higherordergreen]
\tikzstyle{cWire}=[color=quantumgray,thin]
\tikzstyle{rWire}=[color=black,thin]

\tikzstyle{env}=[copoint,regular polygon rotate=0,minimum width=0.2cm, fill=black]

\tikzstyle{probs}=[shape=semicircle,fill=white,draw=black,shape border rotate=180,minimum width=1.2cm]

\tikzstyle{every picture}=[baseline=-0.25em,scale=.5]
\tikzstyle{dotpic}=[] 
\tikzstyle{diredges}=[every to/.style={diredge}]
\tikzstyle{math matrix}=[matrix of math nodes,left delimiter=(,right delimiter=),inner sep=2pt,column sep=1em,row sep=0.5em,nodes={inner sep=0pt},text height=1.5ex, text depth=0.25ex]

\tikzstyle{inline text}=[text height=1.5ex, text depth=0.25ex,yshift=0.5mm]
\tikzstyle{label}=[font=\scriptsize,text height=1.5ex, text depth=0.25ex,yshift=0.0mm]
\tikzstyle{left label}=[label,anchor=east,xshift=1.5mm]
\tikzstyle{right label}=[label,anchor=west,xshift=-1mm]

\tikzstyle{braceedge}=[decorate,decoration={brace,amplitude=2mm,raise=-1mm}]
\tikzstyle{small braceedge}=[decorate,decoration={brace,amplitude=1mm,raise=-1mm}]

\tikzstyle{doubled}=[line width=1.6pt] 
\tikzstyle{boldedge}=[doubled,shorten <=-0.17mm,shorten >=-0.17mm]
\tikzstyle{boldedgegray}=[doubled,gray,shorten <=-0.17mm,shorten >=-0.17mm]
\tikzstyle{singleedgegray}=[gray]

\tikzstyle{semidoubled}=[line width=1.4pt] 
\tikzstyle{semiboldedgegray}=[semidoubled,gray,shorten <=-0.17mm,shorten >=-0.17mm]

\tikzstyle{boxedge}=[semiboldedgegray]

\tikzstyle{boldedgedashed}=[very thick,dashed,shorten <=-0.17mm,shorten >=-0.17mm]
\tikzstyle{vboldedgedashed}=[doubled,dashed,shorten <=-0.17mm,shorten >=-0.17mm]
\tikzstyle{left hook arrow}=[left hook-latex]
\tikzstyle{right hook arrow}=[right hook-latex]
\tikzstyle{sembracket}=[line width=0.5pt,shorten <=-0.07mm,shorten >=-0.07mm]

\tikzstyle{causal edge}=[->,thick,gray]
\tikzstyle{causal nondir}=[thick,gray]
\tikzstyle{timeline}=[thick,gray, dashed]

\tikzstyle{cedge}=[<->,thick,gray!70!white]

\tikzstyle{empty diagram}=[draw=gray!40!white,dashed,shape=rectangle,minimum width=1cm,minimum height=1cm]
\tikzstyle{empty diagram small}=[draw=gray!50!white,dashed,shape=rectangle,minimum width=0.6cm,minimum height=0.5cm]

\tikzstyle{dot}=[inner sep=0mm,minimum width=2mm,minimum height=2mm,draw,shape=circle]
\tikzstyle{leak}=[white dot, shape=regular polygon, minimum size=3.3 mm, regular polygon sides=3, outer sep=-0.2mm, regular polygon rotate=270]
\tikzstyle{proj}=[draw,fill=white,chamfered rectangle,chamfered rectangle angle=30, minimum width=2mm, minimum height= 1mm,scale=0.5, outer sep=-0.2mm]
\tikzstyle{wide proj}=[draw,fill=white,chamfered rectangle,chamfered rectangle angle=30, minimum width=15mm,minimum height=1mm,scale=0.5, outer sep=-0.2mm]
\tikzstyle{very wide proj}=[draw,fill=white,chamfered rectangle,chamfered rectangle angle=30, minimum width=25mm,minimum height=1mm,scale=0.5, outer sep=-0.2mm]
\tikzstyle{very very wide proj}=[draw,fill=white,chamfered rectangle,chamfered rectangle angle=30, minimum width=35mm,minimum height=1mm,scale=0.5, outer sep=-0.2mm]
\tikzstyle{preleak}=[proj]

\tikzstyle{split proj out}=[regular polygon,regular polygon sides=3,draw,scale=0.75,inner sep=-0.5pt,minimum width=3.3mm,fill=white,regular polygon rotate=180]
\tikzstyle{split proj in}=[regular polygon,regular polygon sides=3,draw,scale=0.75,inner sep=-0.5pt,minimum width=3.3mm,fill=white]
\tikzstyle{Vleak}=[white dot, shape=regular polygon, minimum size=3.3 mm, regular polygon sides=3, outer sep=-0.2mm, regular polygon rotate=90]
\tikzstyle{dleak}=[white dot, line width=1.6pt, shape=regular polygon, minimum size=3.3 mm, regular polygon sides=3, outer sep=-0.2mm, regular polygon rotate=270]

\tikzstyle{Wsquare}=[white dot, shape=regular polygon, rounded corners=0.8 mm, minimum size=3.3 mm, regular polygon sides=3, outer sep=-0.2mm]
\tikzstyle{Wsquareadj}=[white dot, shape=regular polygon, rounded corners=0.8 mm, minimum size=3.3 mm, regular polygon sides=3, outer sep=-0.2mm, regular polygon rotate=180]
\tikzstyle{ddot}=[inner sep=0mm, doubled, minimum width=2.5mm,minimum height=2.5mm,draw,shape=circle]

\tikzstyle{black dot}=[dot,fill=black]
\tikzstyle{white dot}=[dot,fill=white,,text depth=-0.2mm]
\tikzstyle{white Wsquare}=[Wsquare,fill=gray,,text depth=-0.2mm]
\tikzstyle{white Wsquareadj}=[Wsquareadj,fill=white,,text depth=-0.2mm]
\tikzstyle{green dot}=[white dot] 
\tikzstyle{gray dot}=[dot,fill=gray!40!white,,text depth=-0.2mm]
\tikzstyle{red dot}=[gray dot] 

\tikzstyle{black ddot}=[ddot,fill=black]
\tikzstyle{white ddot}=[ddot,fill=white]
\tikzstyle{gray ddot}=[ddot,fill=gray!40!white]

\tikzstyle{gray edge}=[gray!60!white]

\tikzstyle{small dot}=[inner sep=0.5mm,minimum width=0pt,minimum height=0pt,draw,shape=circle]

\tikzstyle{small black dot}=[small dot,fill=black]
\tikzstyle{small white dot}=[small dot,fill=white]
\tikzstyle{small gray dot}=[small dot,fill=gray!40!white]

\tikzstyle{causal dot}=[inner sep=0.4mm,minimum width=0pt,minimum height=0pt,draw=white,shape=circle,fill=gray!40!white]

\tikzstyle{phase dimensions}=[minimum size=5mm,font=\footnotesize,rectangle,rounded corners=2.5mm,inner sep=0.2mm,outer sep=-2mm]
\tikzstyle{dphase dimensions}=[minimum size=5mm,font=\footnotesize,rectangle,rounded corners=2.5mm,inner sep=0.2mm,outer sep=-2mm]

\tikzstyle{white phase dot}=[dot,fill=white,phase dimensions]
\tikzstyle{white phase ddot}=[ddot,fill=white,dphase dimensions]

\tikzstyle{white rect ddot}=[draw=black,fill=white,doubled,minimum size=5mm,font=\footnotesize,rectangle,rounded corners=2.5mm,inner sep=0.2mm]
\tikzstyle{gray rect ddot}=[draw=black,fill=gray!40!white,doubled,minimum size=6mm,font=\footnotesize,rectangle,rounded corners=3mm]

\tikzstyle{gray phase dot}=[dot,fill=gray!40!white,phase dimensions]
\tikzstyle{gray phase ddot}=[ddot,fill=gray!40!white,dphase dimensions]
\tikzstyle{grey phase dot}=[gray phase dot]
\tikzstyle{grey phase ddot}=[gray phase ddot]

\tikzstyle{small phase dimensions}=[minimum size=4mm,font=\tiny,rectangle,rounded corners=2mm,inner sep=0.2mm,outer sep=-2mm]
\tikzstyle{small dphase dimensions}=[minimum size=4mm,font=\tiny,rectangle,rounded corners=2mm,inner sep=0.2mm,outer sep=-2mm]

\tikzstyle{small gray phase dot}=[dot,fill=gray!40!white,small phase dimensions]
\tikzstyle{small gray phase ddot}=[ddot,fill=gray!40!white,small dphase dimensions]

\tikzstyle{small map}=[draw,shape=rectangle,minimum height=4mm,minimum width=4mm,fill=white]

\tikzstyle{cnot}=[fill=white,shape=circle,inner sep=-1.4pt]

\tikzstyle{asym hadamard}=[fill=white,draw,shape=NEbox,inner sep=0.6mm,font=\footnotesize,minimum height=4mm]
\tikzstyle{asym hadamard conj}=[fill=white,draw,shape=NWbox,inner sep=0.6mm,font=\footnotesize,minimum height=4mm]
\tikzstyle{asym hadamard dag}=[fill=white,draw,shape=SEbox,inner sep=0.6mm,font=\footnotesize,minimum height=4mm]

\tikzstyle{hadamard}=[fill=white,draw,inner sep=0.6mm,font=\footnotesize,minimum height=4mm,minimum width=4mm]
\tikzstyle{small hadamard}=[fill=white,draw,inner sep=0.6mm,minimum height=1.5mm,minimum width=1.5mm]
\tikzstyle{small hadamard rotate}=[small hadamard,rotate=45]
\tikzstyle{dhadamard}=[hadamard,doubled]
\tikzstyle{small dhadamard}=[small hadamard,doubled]
\tikzstyle{small dhadamard rotate}=[small hadamard rotate,doubled]
\tikzstyle{antipode}=[white dot,inner sep=0.3mm,font=\footnotesize]

\tikzstyle{scalar}=[diamond,draw,inner sep=0.5pt,font=\small]
\tikzstyle{dscalar}=[diamond,doubled, draw,inner sep=0.5pt,font=\small]

\tikzstyle{small box}=[rectangle,inline text,fill=white,draw,minimum height=5mm,yshift=-0.5mm,minimum width=5mm,font=\small]
\tikzstyle{small gray box}=[small box,fill=gray!30]
\tikzstyle{medium box}=[rectangle,inline text,fill=white,draw,minimum height=5mm,yshift=-0.5mm,minimum width=10mm,font=\small]
\tikzstyle{square box}=[small box] 
\tikzstyle{medium gray box}=[small box,fill=gray!30]
\tikzstyle{semilarge box}=[rectangle,inline text,fill=white,draw,minimum height=5mm,yshift=-0.5mm,minimum width=12.5mm,font=\small]
\tikzstyle{large box}=[rectangle,inline text,fill=white,draw,minimum height=5mm,yshift=-0.5mm,minimum width=15mm,font=\small]
\tikzstyle{large gray box}=[small box,fill=gray!30]

\tikzstyle{Bayes box}=[rectangle,fill=black,draw, minimum height=3mm, minimum width=3mm]

\tikzstyle{gray square point}=[small box,fill=gray!50]

\tikzstyle{dphase box white}=[dhadamard]
\tikzstyle{dphase box gray}=[dhadamard,fill=gray!50!white]
\tikzstyle{phase box white}=[hadamard]
\tikzstyle{phase box gray}=[hadamard,fill=gray!50!white]

\tikzstyle{point}=[regular polygon,regular polygon sides=3,draw,scale=0.75,inner sep=-0.5pt,minimum width=9mm,fill=white,regular polygon rotate=180]
\tikzstyle{point nosep}=[regular polygon,regular polygon sides=3,draw,scale=0.75,inner sep=-2pt,minimum width=9mm,fill=white,regular polygon rotate=180]
\tikzstyle{copoint}=[regular polygon,regular polygon sides=3,draw,scale=0.75,inner sep=-0.5pt,minimum width=9mm,fill=white]
\tikzstyle{dpoint}=[point,doubled]
\tikzstyle{dcopoint}=[copoint,doubled]

\tikzstyle{pointgrow}=[shape=cornerpoint,kpoint common,scale=0.75,inner sep=3pt]
\tikzstyle{pointgrow dag}=[shape=cornercopoint,kpoint common,scale=0.75,inner sep=3pt]

\tikzstyle{wide copoint}=[fill=white,draw,shape=isosceles triangle,shape border rotate=90,isosceles triangle stretches=true,inner sep=0pt,minimum width=1.5cm,minimum height=6.12mm]
\tikzstyle{wide point}=[fill=white,draw,shape=isosceles triangle,shape border rotate=-90,isosceles triangle stretches=true,inner sep=0pt,minimum width=1.5cm,minimum height=6.12mm,yshift=-0.0mm]
\tikzstyle{wide point plus}=[fill=white,draw,shape=isosceles triangle,shape border rotate=-90,isosceles triangle stretches=true,inner sep=0pt,minimum width=1.74cm,minimum height=7mm,yshift=-0.0mm]

\tikzstyle{wide dpoint}=[fill=white,doubled,draw,shape=isosceles triangle,shape border rotate=-90,isosceles triangle stretches=true,inner sep=0pt,minimum width=1.5cm,minimum height=6.12mm,yshift=-0.0mm]

\tikzstyle{tinypoint}=[regular polygon,regular polygon sides=3,draw,scale=0.55,inner sep=-0.15pt,minimum width=6mm,fill=white,regular polygon rotate=180]

\tikzstyle{white point}=[point]
\tikzstyle{white dpoint}=[dpoint]
\tikzstyle{green point}=[white point] 
\tikzstyle{white copoint}=[copoint]
\tikzstyle{gray point}=[point,fill=gray!40!white]
\tikzstyle{gray dpoint}=[gray point,doubled]
\tikzstyle{red point}=[gray point] 
\tikzstyle{gray copoint}=[copoint,fill=gray!40!white]
\tikzstyle{gray dcopoint}=[gray copoint,doubled]

\tikzstyle{white point guide}=[regular polygon,regular polygon sides=3,font=\scriptsize,draw,scale=0.65,inner sep=-0.5pt,minimum width=9mm,fill=white,regular polygon rotate=180]

\tikzstyle{black point}=[point,fill=black,font=\color{white}]
\tikzstyle{black copoint}=[copoint,fill=black,font=\color{white}]

\tikzstyle{tiny gray point}=[tinypoint,fill=gray!40!white]

\tikzstyle{diredge}=[->]
\tikzstyle{ddiredge}=[<->]
\tikzstyle{rdiredge}=[<-]
\tikzstyle{thickdiredge}=[->, very thick]
\tikzstyle{pointer edge}=[->,very thick,gray]
\tikzstyle{pointer edge part}=[very thick,gray]
\tikzstyle{dashed edge}=[dashed]
\tikzstyle{thick dashed edge}=[very thick,dashed]
\tikzstyle{thick gray dashed edge}=[thick dashed edge,gray!40]
\tikzstyle{thick map edge}=[very thick,|->]

\makeatletter
\newcommand{\boxshape}[3]{%
\pgfdeclareshape{#1}{
\inheritsavedanchors[from=rectangle] 
\inheritanchorborder[from=rectangle]
\inheritanchor[from=rectangle]{center}
\inheritanchor[from=rectangle]{north}
\inheritanchor[from=rectangle]{south}
\inheritanchor[from=rectangle]{west}
\inheritanchor[from=rectangle]{east}
\backgroundpath{
\southwest \pgf@xa=\pgf@x \pgf@ya=\pgf@y
\northeast \pgf@xb=\pgf@x \pgf@yb=\pgf@y

\@tempdima=#2
\@tempdimb=#3

\pgfpathmoveto{\pgfpoint{\pgf@xa - 5pt + \@tempdima}{\pgf@ya}}
\pgfpathlineto{\pgfpoint{\pgf@xa - 5pt - \@tempdima}{\pgf@yb}}
\pgfpathlineto{\pgfpoint{\pgf@xb + 5pt + \@tempdimb}{\pgf@yb}}
\pgfpathlineto{\pgfpoint{\pgf@xb + 5pt - \@tempdimb}{\pgf@ya}}
\pgfpathlineto{\pgfpoint{\pgf@xa - 5pt + \@tempdima}{\pgf@ya}}
\pgfpathclose
}
}}

\boxshape{NEbox}{0pt}{3pt}
\boxshape{SEbox}{0pt}{-3pt}
\boxshape{NWbox}{5pt}{0pt}
\boxshape{SWbox}{-5pt}{0pt}
\boxshape{EBox}{-3pt}{3pt}
\boxshape{WBox}{3pt}{-3pt}
\makeatother

\tikzstyle{cloud}=[shape=cloud,draw,minimum width=1.5cm,minimum height=1.5cm]

\tikzstyle{map}=[draw,shape=NEbox,inner sep=1pt,minimum height=4mm,fill=white]
\tikzstyle{dashedmap}=[draw,dashed,shape=NEbox,inner sep=2pt,minimum height=6mm,fill=white]
\tikzstyle{mapdag}=[draw,shape=SEbox,inner sep=1pt,minimum height=4mm,fill=white]
\tikzstyle{mapadj}=[draw,shape=SEbox,inner sep=2pt,minimum height=6mm,fill=white]
\tikzstyle{maptrans}=[draw,shape=SWbox,inner sep=2pt,minimum height=6mm,fill=white]
\tikzstyle{mapconj}=[draw,shape=NWbox,inner sep=2pt,minimum height=6mm,fill=white]

\tikzstyle{medium map}=[draw,shape=NEbox,inner sep=2pt,minimum height=6mm,fill=white,minimum width=7mm]
\tikzstyle{medium map dag}=[draw,shape=SEbox,inner sep=2pt,minimum height=6mm,fill=white,minimum width=7mm]
\tikzstyle{medium map adj}=[draw,shape=SEbox,inner sep=2pt,minimum height=6mm,fill=white,minimum width=7mm]
\tikzstyle{medium map trans}=[draw,shape=SWbox,inner sep=2pt,minimum height=6mm,fill=white,minimum width=7mm]
\tikzstyle{medium map conj}=[draw,shape=NWbox,inner sep=2pt,minimum height=6mm,fill=white,minimum width=7mm]
\tikzstyle{semilarge map}=[draw,shape=NEbox,inner sep=2pt,minimum height=6mm,fill=white,minimum width=9.5mm]
\tikzstyle{semilarge map trans}=[draw,shape=SWbox,inner sep=2pt,minimum height=6mm,fill=white,minimum width=9.5mm]
\tikzstyle{semilarge map adj}=[draw,shape=SEbox,inner sep=2pt,minimum height=6mm,fill=white,minimum width=9.5mm]
\tikzstyle{semilarge map dag}=[draw,shape=SEbox,inner sep=2pt,minimum height=6mm,fill=white,minimum width=9.5mm]
\tikzstyle{semilarge map conj}=[draw,shape=NWbox,inner sep=2pt,minimum height=6mm,fill=white,minimum width=9.5mm]
\tikzstyle{large map}=[draw,shape=NEbox,inner sep=2pt,minimum height=6mm,fill=white,minimum width=12mm]
\tikzstyle{large map conj}=[draw,shape=NWbox,inner sep=2pt,minimum height=6mm,fill=white,minimum width=12mm]
\tikzstyle{very large map}=[draw,shape=NEbox,inner sep=2pt,minimum height=6mm,fill=white,minimum width=17mm]

\tikzstyle{medium dmap}=[draw,doubled,shape=NEbox,inner sep=2pt,minimum height=6mm,fill=white,minimum width=7mm]
\tikzstyle{medium dmap dag}=[draw,doubled,shape=SEbox,inner sep=2pt,minimum height=6mm,fill=white,minimum width=7mm]
\tikzstyle{medium dmap adj}=[draw,doubled,shape=SEbox,inner sep=2pt,minimum height=6mm,fill=white,minimum width=7mm]
\tikzstyle{medium dmap trans}=[draw,doubled,shape=SWbox,inner sep=2pt,minimum height=6mm,fill=white,minimum width=7mm]
\tikzstyle{medium dmap conj}=[draw,doubled,shape=NWbox,inner sep=2pt,minimum height=6mm,fill=white,minimum width=7mm]
\tikzstyle{semilarge dmap}=[draw,doubled,shape=NEbox,inner sep=2pt,minimum height=6mm,fill=white,minimum width=9.5mm]
\tikzstyle{semilarge dmap trans}=[draw,doubled,shape=SWbox,inner sep=2pt,minimum height=6mm,fill=white,minimum width=9.5mm]
\tikzstyle{semilarge dmap adj}=[draw,doubled,shape=SEbox,inner sep=2pt,minimum height=6mm,fill=white,minimum width=9.5mm]
\tikzstyle{semilarge dmap dag}=[draw,doubled,shape=SEbox,inner sep=2pt,minimum height=6mm,fill=white,minimum width=9.5mm]
\tikzstyle{semilarge dmap conj}=[draw,doubled,shape=NWbox,inner sep=2pt,minimum height=6mm,fill=white,minimum width=9.5mm]
\tikzstyle{large dmap}=[draw,doubled,shape=NEbox,inner sep=2pt,minimum height=6mm,fill=white,minimum width=12mm]
\tikzstyle{large dmap conj}=[draw,doubled,shape=NWbox,inner sep=2pt,minimum height=6mm,fill=white,minimum width=12mm]
\tikzstyle{large dmap trans}=[draw,doubled,shape=SWbox,inner sep=2pt,minimum height=6mm,fill=white,minimum width=12mm]
\tikzstyle{large dmap adj}=[draw,doubled,shape=SEbox,inner sep=2pt,minimum height=6mm,fill=white,minimum width=12mm]
\tikzstyle{large dmap dag}=[draw,doubled,shape=SEbox,inner sep=2pt,minimum height=6mm,fill=white,minimum width=12mm]
\tikzstyle{very large dmap}=[draw,doubled,shape=NEbox,inner sep=2pt,minimum height=6mm,fill=white,minimum width=19.5mm]

\tikzstyle{muxbox}=[draw,shape=rectangle,minimum height=3mm,minimum width=3mm,fill=white]
\tikzstyle{dmuxbox}=[muxbox,doubled]

\tikzstyle{box}=[draw,shape=rectangle,inner sep=2pt,minimum height=6mm,minimum width=6mm,fill=white]
\tikzstyle{dbox}=[draw,doubled,shape=rectangle,inner sep=2pt,minimum height=6mm,minimum width=6mm,fill=white]
\tikzstyle{dmap}=[draw,doubled,shape=NEbox,inner sep=2pt,minimum height=6mm,fill=white]
\tikzstyle{dmapdag}=[draw,doubled,shape=SEbox,inner sep=2pt,minimum height=6mm,fill=white]
\tikzstyle{dmapadj}=[draw,doubled,shape=SEbox,inner sep=2pt,minimum height=6mm,fill=white]
\tikzstyle{dmaptrans}=[draw,doubled,shape=SWbox,inner sep=2pt,minimum height=6mm,fill=white]
\tikzstyle{dmapconj}=[draw,doubled,shape=NWbox,inner sep=2pt,minimum height=6mm,fill=white]

\tikzstyle{ddmap}=[draw,doubled,dashed,shape=NEbox,inner sep=2pt,minimum height=6mm,fill=white]
\tikzstyle{ddmapdag}=[draw,doubled,dashed,shape=SEbox,inner sep=2pt,minimum height=6mm,fill=white]
\tikzstyle{ddmapadj}=[draw,doubled,dashed,shape=SEbox,inner sep=2pt,minimum height=6mm,fill=white]
\tikzstyle{ddmaptrans}=[draw,doubled,dashed,shape=SWbox,inner sep=2pt,minimum height=6mm,fill=white]
\tikzstyle{ddmapconj}=[draw,doubled,dashed,shape=NWbox,inner sep=2pt,minimum height=6mm,fill=white]

\boxshape{sNEbox}{0pt}{3pt}
\boxshape{sSEbox}{0pt}{-3pt}
\boxshape{sNWbox}{3pt}{0pt}
\boxshape{sSWbox}{-3pt}{0pt}
\tikzstyle{smap}=[draw,shape=sNEbox,fill=white]
\tikzstyle{smapdag}=[draw,shape=sSEbox,fill=white]
\tikzstyle{smapadj}=[draw,shape=sSEbox,fill=white]
\tikzstyle{smaptrans}=[draw,shape=sSWbox,fill=white]
\tikzstyle{smapconj}=[draw,shape=sNWbox,fill=white]

\tikzstyle{dsmap}=[draw,dashed,shape=sNEbox,fill=white]
\tikzstyle{dsmapdag}=[draw,dashed,shape=sSEbox,fill=white]
\tikzstyle{dsmaptrans}=[draw,dashed,shape=sSWbox,fill=white]
\tikzstyle{dsmapconj}=[draw,dashed,shape=sNWbox,fill=white]

\boxshape{mNEbox}{0pt}{10pt}
\boxshape{mSEbox}{0pt}{-10pt}
\boxshape{mNWbox}{10pt}{0pt}
\boxshape{mSWbox}{-10pt}{0pt}
\tikzstyle{mmap}=[draw,shape=mNEbox]
\tikzstyle{mmapdag}=[draw,shape=mSEbox]
\tikzstyle{mmaptrans}=[draw,shape=mSWbox]
\tikzstyle{mmapconj}=[draw,shape=mNWbox]

\tikzstyle{mmapgray}=[draw,fill=gray!40!white,shape=mNEbox]
\tikzstyle{smapgray}=[draw,fill=gray!40!white,shape=sNEbox]

\makeatletter

\pgfdeclareshape{cornerpoint}{
\inheritsavedanchors[from=rectangle] 
\inheritanchorborder[from=rectangle]
\inheritanchor[from=rectangle]{center}
\inheritanchor[from=rectangle]{north}
\inheritanchor[from=rectangle]{south}
\inheritanchor[from=rectangle]{west}
\inheritanchor[from=rectangle]{east}
\backgroundpath{
\southwest \pgf@xa=\pgf@x \pgf@ya=\pgf@y
\northeast \pgf@xb=\pgf@x \pgf@yb=\pgf@y

\pgfmathsetmacro{\pgf@shorten@left}{\pgfkeysvalueof{/tikz/shorten left}}
\pgfmathsetmacro{\pgf@shorten@right}{\pgfkeysvalueof{/tikz/shorten right}}

\pgfpathmoveto{\pgfpoint{0.5 * (\pgf@xa + \pgf@xb)}{\pgf@ya - 5pt}}
\pgfpathlineto{\pgfpoint{\pgf@xa - 8pt + \pgf@shorten@left}{\pgf@yb - 1.5 * \pgf@shorten@left}}
\pgfpathlineto{\pgfpoint{\pgf@xa - 8pt + \pgf@shorten@left}{\pgf@yb}}
\pgfpathlineto{\pgfpoint{\pgf@xb + 8pt - \pgf@shorten@right}{\pgf@yb}}
\pgfpathlineto{\pgfpoint{\pgf@xb + 8pt - \pgf@shorten@right}{\pgf@yb - 1.5 * \pgf@shorten@right}}
\pgfpathclose
}
}

\pgfdeclareshape{cornercopoint}{
\inheritsavedanchors[from=rectangle] 
\inheritanchorborder[from=rectangle]
\inheritanchor[from=rectangle]{center}
\inheritanchor[from=rectangle]{north}
\inheritanchor[from=rectangle]{south}
\inheritanchor[from=rectangle]{west}
\inheritanchor[from=rectangle]{east}
\backgroundpath{
\southwest \pgf@xa=\pgf@x \pgf@ya=\pgf@y
\northeast \pgf@xb=\pgf@x \pgf@yb=\pgf@y

\pgfmathsetmacro{\pgf@shorten@left}{\pgfkeysvalueof{/tikz/shorten left}}
\pgfmathsetmacro{\pgf@shorten@right}{\pgfkeysvalueof{/tikz/shorten right}}

\pgfpathmoveto{\pgfpoint{0.5 * (\pgf@xa + \pgf@xb)}{\pgf@yb + 5pt}}
\pgfpathlineto{\pgfpoint{\pgf@xa - 8pt + \pgf@shorten@left}{\pgf@ya + 1.5 * \pgf@shorten@left}}
\pgfpathlineto{\pgfpoint{\pgf@xa - 8pt + \pgf@shorten@left}{\pgf@ya}}
\pgfpathlineto{\pgfpoint{\pgf@xb + 8pt - \pgf@shorten@right}{\pgf@ya}}
\pgfpathlineto{\pgfpoint{\pgf@xb + 8pt - \pgf@shorten@right}{\pgf@ya + 1.5 * \pgf@shorten@right}}
\pgfpathclose
}
}

\makeatother

\pgfkeyssetvalue{/tikz/shorten left}{0pt}
\pgfkeyssetvalue{/tikz/shorten right}{0pt}

\tikzstyle{kpoint common}=[draw,fill=white,inner sep=1pt,minimum height=4mm]
\tikzstyle{kpoint sc}=[shape=cornerpoint,kpoint common]
\tikzstyle{kpoint adjoint sc}=[shape=cornercopoint,kpoint common]
\tikzstyle{kpoint}=[shape=cornerpoint,shorten left=5pt,kpoint common]
\tikzstyle{kpoint adjoint}=[shape=cornercopoint,shorten left=5pt,kpoint common]
\tikzstyle{kpoint conjugate}=[shape=cornerpoint,shorten right=5pt,kpoint common]
\tikzstyle{kpoint transpose}=[shape=cornercopoint,shorten right=5pt,kpoint common]
\tikzstyle{kpoint symm}=[shape=cornerpoint,shorten left=5pt,shorten right=5pt,kpoint common]

\tikzstyle{wide kpoint sc}=[shape=cornerpoint,kpoint common, minimum width=1 cm]
\tikzstyle{wide kpointdag sc}=[shape=cornercopoint,kpoint common, minimum width=1 cm]

\tikzstyle{black kpoint}=[shape=cornerpoint,shorten left=5pt,kpoint common,fill=black,font=\color{white}]

\tikzstyle{black kpoint sm}=[shape=cornerpoint,shorten left=5pt,kpoint common,fill=black,font=\color{white},scale=0.75]

\tikzstyle{black kpoint adjoint}=[shape=cornercopoint,shorten left=5pt,kpoint common,fill=black,font=\color{white}]
\tikzstyle{black kpointadj}=[shape=cornercopoint,shorten left=5pt,kpoint common,fill=black,font=\color{white}]

\tikzstyle{black kpointadj sm}=[shape=cornercopoint,shorten left=5pt,kpoint common,fill=black,font=\color{white},scale=0.75]

\tikzstyle{black dkpoint}=[shape=cornerpoint,shorten left=5pt,kpoint common,fill=black, doubled,font=\color{white}]
\tikzstyle{black dkpoint adjoint}=[shape=cornercopoint,shorten left=5pt,kpoint common,fill=black, doubled,font=\color{white}]
\tikzstyle{black dkpointadj}=[shape=cornercopoint,shorten left=5pt,kpoint common,fill=black, doubled,font=\color{white}]

\tikzstyle{black dkpoint sm}=[shape=cornerpoint,shorten left=5pt,kpoint common,fill=black, doubled,font=\color{white},scale=0.75]
\tikzstyle{black dkpointadj sm}=[shape=cornercopoint,shorten left=5pt,kpoint common,fill=black, doubled,font=\color{white},scale=0.75]

\tikzstyle{kpointdag}=[kpoint adjoint]
\tikzstyle{kpointadj}=[kpoint adjoint]
\tikzstyle{kpointconj}=[kpoint conjugate]
\tikzstyle{kpointtrans}=[kpoint transpose]

\tikzstyle{big kpoint}=[kpoint, minimum width=1.2 cm, minimum height=8mm, inner sep=4pt, text depth=3mm]

\tikzstyle{wide kpoint}=[kpoint, minimum width=1 cm, inner sep=2pt]
\tikzstyle{wide kpointdag}=[kpointdag, minimum width=1 cm, inner sep=2pt]
\tikzstyle{wide kpointconj}=[kpointconj, minimum width=1 cm, inner sep=2pt]
\tikzstyle{wide kpointtrans}=[kpointtrans, minimum width=1 cm, inner sep=2pt]

\tikzstyle{wider kpoint}=[kpoint, minimum width=1.25 cm, inner sep=2pt]
\tikzstyle{wider kpointdag}=[kpointdag, minimum width=1.25 cm, inner sep=2pt]
\tikzstyle{wider kpointconj}=[kpointconj, minimum width=1.25 cm, inner sep=2pt]
\tikzstyle{wider kpointtrans}=[kpointtrans, minimum width=1.25 cm, inner sep=2pt]

\tikzstyle{gray kpoint}=[kpoint,fill=gray!50!white]
\tikzstyle{gray kpointdag}=[kpointdag,fill=gray!50!white]
\tikzstyle{gray kpointadj}=[kpointadj,fill=gray!50!white]
\tikzstyle{gray kpointconj}=[kpointconj,fill=gray!50!white]
\tikzstyle{gray kpointtrans}=[kpointtrans,fill=gray!50!white]

\tikzstyle{gray dkpoint}=[kpoint,fill=gray!50!white,doubled]
\tikzstyle{gray dkpointdag}=[kpointdag,fill=gray!50!white,doubled]
\tikzstyle{gray dkpointadj}=[kpointadj,fill=gray!50!white,doubled]
\tikzstyle{gray dkpointconj}=[kpointconj,fill=gray!50!white,doubled]
\tikzstyle{gray dkpointtrans}=[kpointtrans,fill=gray!50!white,doubled]

\tikzstyle{white label}=[draw,fill=white,rectangle,inner sep=0.7 mm]
\tikzstyle{gray label}=[draw,fill=gray!50!white,rectangle,inner sep=0.7 mm]
\tikzstyle{black label}=[draw,fill=black,rectangle,inner sep=0.7 mm]

\tikzstyle{dkpoint}=[kpoint,doubled]
\tikzstyle{wide dkpoint}=[wide kpoint,doubled]
\tikzstyle{dkpointdag}=[kpoint adjoint,doubled]
\tikzstyle{wide dkpointdag}=[wide kpointdag,doubled]
\tikzstyle{dkcopoint}=[kpoint adjoint,doubled]
\tikzstyle{dkpointadj}=[kpoint adjoint,doubled]
\tikzstyle{dkpointconj}=[kpoint conjugate,doubled]
\tikzstyle{dkpointtrans}=[kpoint transpose,doubled]

\tikzstyle{kscalar}=[kpoint common, shape=EBox, inner xsep=-1pt, inner ysep=3pt,font=\small]
\tikzstyle{kscalarconj}=[kpoint common, shape=WBox, inner xsep=-1pt, inner ysep=3pt,font=\small]

\tikzstyle{spekpoint}=[kpoint sc,minimum height=5mm,inner sep=3pt]
\tikzstyle{spekcopoint}=[kpoint adjoint sc,minimum height=5mm,inner sep=3pt]

\tikzstyle{dspekpoint}=[spekpoint,doubled]
\tikzstyle{dspekcopoint}=[spekcopoint,doubled]

 \tikzstyle{upground}=[circuit ee IEC,thick,ground,rotate=90,scale=2.5]
 \tikzstyle{downground}=[circuit ee IEC,thick,ground,rotate=-90,scale=2.5]
 \tikzstyle{bigground}=[regular polygon,regular polygon sides=3,draw=gray,scale=0.50,inner sep=-0.5pt,minimum width=10mm,fill=gray]

\tikzstyle{arrs}=[-latex,font=\small,auto]
\tikzstyle{arrow plain}=[arrs]
\tikzstyle{arrow dashed}=[dashed,arrs]
\tikzstyle{arrow bold}=[very thick,arrs]
\tikzstyle{arrow hide}=[draw=white!0,-]
\tikzstyle{arrow reverse}=[latex-]
\tikzstyle{cdnode}=[]

\newcommand{\bit}{\begin{itemize}}
\newcommand{\eit}{\end{itemize}\par\noindent}
\newcommand{\ben}{\begin{enumerate}}
\newcommand{\een}{\end{enumerate}\par\noindent}
\newcommand{\beqa}{\begin{eqnarray*}}
\newcommand{\eeqa}{\end{eqnarray*}\par\noindent}
\newcommand{\beqn}{\begin{eqnarray}}
\newcommand{\eeqn}{\end{eqnarray}\par\noindent}

\definecolor{quantumviolet}{RGB}{79, 4, 134}
\definecolor{quantumgray}{RGB}{134,79,4}
\definecolor{higherorderred}{RGB}{134,0,0}
\definecolor{higherordergreen}{RGB}{0,134,0}

\pgfplotsset{compat=newest}

\usetikzlibrary{positioning,backgrounds}
\usetikzlibrary{calc,intersections,through}

\usepackage{microtype}
\usepackage[all=normal,floats=tight,mathdisplays=normal,mathspacing=tight,wordspacing=tight,paragraphs=normal,tracking=tight,charwidths=tight,sections=normal,margins=normal]{savetrees}
\everypar=\expandafter{\the\everypar\loosness=-1 }
\begin{document}

 \title{Contextuality without incompatibility}
\author{John H. Selby}
\email{john.h.selby@gmail.com}
\affiliation{International Centre for Theory of Quantum Technologies, University of Gda\'nsk, 80-309 Gda\'nsk, Poland}
\author{David Schmid}
\affiliation{International Centre for Theory of Quantum Technologies, University of Gda\'nsk, 80-309 Gda\'nsk, Poland}
\affiliation{Perimeter Institute for Theoretical Physics, 31 Caroline Street North, Waterloo, Ontario Canada N2L 2Y5}
\affiliation{Institute for Quantum Computing and Department of Physics and Astronomy, University of Waterloo, Waterloo, Ontario N2L 3G1, Canada}
\author{Elie Wolfe}
\affiliation{Perimeter Institute for Theoretical Physics, 31 Caroline Street North, Waterloo, Ontario Canada N2L 2Y5}
\author{Ana Bel\'en Sainz}
\affiliation{International Centre for Theory of Quantum Technologies, University of Gda\'nsk, 80-309 Gda\'nsk, Poland}
\author{Ravi Kunjwal}
\affiliation{Centre for Quantum Information and Communication, Ecole polytechnique de Bruxelles,
	CP 165, Universit\'e libre de Bruxelles, 1050 Brussels, Belgium}
\author{Robert W. Spekkens}
\affiliation{Perimeter Institute for Theoretical Physics, 31 Caroline Street North, Waterloo, Ontario Canada N2L 2Y5}

\begin{abstract}
The existence of incompatible measurements is often believed to be a feature of quantum theory which signals its inconsistency with any classical worldview.
To prove the failure of classicality in the sense of Kochen-Specker noncontextuality, one does indeed require sets of incompatible measurements. 
However, a more broadly applicable 
notion of classicality is the existence of a generalized-noncontextual ontological model. 
In particular, this notion can imply constraints on the representation of outcomes even within a single nonprojective measurement.  
We leverage this fact to demonstrate that measurement incompatibility is neither necessary nor sufficient for proofs of the failure of generalized noncontextuality. Furthermore, we show that every proof of the failure of generalized noncontextuality in a quantum prepare-measure scenario can be converted into a proof of the failure of generalized noncontextuality in a corresponding scenario with no incompatible measurements.
\end{abstract}

\maketitle

Measurement incompatibility---the existence of measurements that cannot be implemented simultaneously---has conventionally been taken to be part of what is truly distinctive about quantum theory relative to its classical forebears.  We are here concerned with whether this attitude is justified when the notion of classicality at play is whether or not a theory (or experiment) admits of an ontological model satisfying the principle of generalized noncontextuality~\cite{gencontext,schmid2020unscrambling}.
It is already known that classical statistical theories with an epistemic restriction~\cite{epistricted}  (and subtheories of quantum theory which make the same predictions as these) can manifest incompatibility despite satisfying the principle of generalized noncontextuality. Hence, the mere fact that a theory exhibits measurement incompatibility is not sufficient to infer that it must fail to admit of a generalized-noncontextual ontological model.

In this article, we demonstrate that measurement incompatibility is also not {\em necessary} for  demonstrating that quantum theory (or an experiment within quantum theory) is  nonclassical in this sense.
In summary, we have
\begin{equation*}
	    \begin{tikzpicture}[scale=1]
		\node at (-4.5, 0.3) {Incompatibility};
		\node at (5.5, 1) {Impossibility of a};
		\node at (5.5, 0.3) {generalized-noncontextual};
		\node at (5.5, -0.4) {ontological model};
\node[scale=1.2] at (0,1) {$\overset{\text{\cite{gencontext,schmid2020unscrambling}}}{\not\Rightarrow}$};
\node[scale=1.2] at (0,-0.2) {$\underset{\text{Thm.~1}}{\not\Leftarrow}$};		
	    \end{tikzpicture}	    
\end{equation*}

In a companion paper, Ref.~\cite{companionincompatibility}, we give more general arguments and applications of these results  (see the conclusions for a summary).

The notion of a generalized-noncontextual
 ontological model, introduced in Ref.~\cite{gencontext}, was proposed to overcome some of the limitations  of the Kochen-Specker notion of noncontextuality, and has since been further refined~\cite{schmid2020unscrambling}.   
It can be understood as a special case of a methodological principle due to Leibniz, a version of the principle of the identity of indiscernibles~\cite{Leibniz,schmid2020unscrambling}.
Unlike the Kochen-Specker notion, generalized noncontextuality 
has implications not only for sharp measurements but for all procedures, including unsharp measurements, preparations, and transformations.
It has been found to subsume many other notions of classicality~\cite{negativity,SchmidGPT,ShahandehGPT,schmid2020structure,baldijao2021noncontextuality} and to shed light on the resource of nonclassicality underlying the quantum
 advantages known to exist for many information processing tasks~\cite{POM,RAC,RAC2,Saha_2019,comp1,comp2,schmid2021only,schmid2018contextual,cloningcontext,contextmetrology,AWV,KLP19,YK20}.

Recall that the notion of Kochen-Specker noncontextuality is defined only for projective measurements (i.e., those whose outcomes correspond to the eigenspaces of a Hermitian operator), and the 
context-independence of the ontological representation of a measurement
 is understood as the lack of dependence on what other measurement 
is implemented simultaneously with it.
It is well-known that the notion of Kochen-Specker noncontextuality only implies nontrivial constraints on the ontological representation if the set of measurements under consideration includes some incompatible ones.

By contrast, the notion of generalized noncontextuality provides a much broader scope of possibilities for assuming context-independence: namely, any two procedures which are operationally equivalent~\cite{gencontext} are assumed to have the same ontological representation.
 As it turns out, for unsharp measurements---which in quantum theory are associated with a positive operator-valued measure (POVM) that is not projector-valued ---there can be nontrivial operational equivalences between the outcomes of one and the same measurement, so that generalized noncontextuality has nontrivial implications for the ontological representation of even a {\em single} such measurement.   
For instance, consider the POVM $\{ \tfrac{1}{2}\mathbb{1} , \tfrac{1}{2}\mathbb{1}\}$ where $\mathbb{1}$ is the identity operator.  The two outcomes of this measurement are equally likely on all states, and hence its two effects are operationally equivalent.  By generalized noncontextuality, then, they must be  represented in the ontological model by the {\em same}  conditional probability distribution~\cite{determinism}. 

It is the existence of such nontrivial constraints on the ontological representation of a single measurement that opens up the possibility of a proof of the failure of generalized noncontextuality without incompatibility. Note that the possibility is open {\em only if} the measurement is unsharp---any experiment involving a single projective measurement trivially admits of a generalized-noncontextual model. 
We demonstrate here that this possibility is in fact realized, and it is realized in the simplest operational scenario, namely, a prepare-measure experiment. 

Our main result can be summarized as follows:
\begin{theorem}\label{thm1}
From every  quantum  proof of the failure of generalized noncontextuality in a prepare-measure scenario involving incompatible measurements one can construct a proof  that does not require any incompatibility.
\end{theorem}

In the following, the term ``noncontextual'' will be used as a shorthand for ``generalized-noncontextual''. 
We will also use the expression ``proof of contextuality'' as a shorthand for ``proof of the impossibility of a generalized-noncontextual ontological model''  (even though the lesson of such proofs may well be to reject the framework of ontological models while maintaining the Leibnizian methodological principle that underlies noncontextuality~\cite{schmid2020unscrambling}). 

Ref.~\cite{PhysRevResearch.2.013011} makes a claim which is superficially contrary to ours, namely, that incompatibility is necessary and sufficient for proving generalized contextuality.  The claim of {\em sufficiency} in Ref.~\cite{PhysRevResearch.2.013011} is only established relative to the strong assumption that the set of states under consideration includes {\em all} quantum states. This assumption is violated in any real experiment
 and also  by  
 many interesting  subtheories of quantum theory~\cite{epistricted,gottesman1997stabilizer,RevModPhys.84.621}.
The claim of the {\em necessity} of incompatibility in Ref.~\cite{PhysRevResearch.2.013011} is predicated on assuming noncontextuality for preparations alone rather than for preparations and measurements. 

However, the motivation for  assuming noncontextuality for preparations (namely, the Leibnizian methodological principle~\cite{Leibniz,schmid2020unscrambling}) is also a motivation for assuming it for measurements, and consequently it is unnatural to restrict the scope of the assumption to preparations alone.

{\bf Conventional no-go theorems for generalized noncontextuality---}
We now introduce the requisite preliminaries by describing a conventional no-go theorem for generalized noncontextuality in the usual setting of prepare-measure scenarios on a given quantum system $\mathcal{H}$.
Such a scenario is depicted in Fig.~\ref{fig1}(a). The scenario is characterized by a set of quantum states $\{\rho_s\}_s$, indexed by the preparation setting $s\in S$, 
 and a set of POVMs, $\{\{E_{y|t}\}_y\}_t$, indexed by the measurement setting $t\in T$, with elements (termed \textit{effects}) in a given measurement indexed by outcomes $y\in Y$.
  (Note that taking the set $Y$ to be the same cardinality for all values $t \in T$ involves no loss of generality.)
 The observable statistics in this scenario are given by the Born rule, i.e.,
\beq
P^{\rm (q)}(y|s,t)=  \mathsf{tr}(E_{y|t} \rho_s).
\eeq

The features of an operational scenario which drive any proof of contextuality are the  set  of operational equivalences that hold among the preparations, indexed by $ a\in O_{P}$,
and the set of operational equivalences that hold among the measurements, indexed by $b\in O_{M}$. 
 These can be expressed \cite{Schmid2018} as linear constraints on the corresponding states and effects:
\begin{align}\label{opequivs}
&\sum_s \alpha_s^{(a)} \rho_s = 0,\
&\sum_{y,t} \beta_{yt}^{(b)} E_{y|t}=0,
\end{align}
 for all $a \in O_P$  and $b \in O_M$  where $\alpha_s^{(a)}$ and $\beta_{yt}^{(b)}$ are real coefficients which specify the operational equivalence.  

An ontological model of the prepare-measure scenario associates an ontic state space $\Lambda$ with the system $\mathcal{H}$ and explains the 
operational statistics as arising from stochastic processes on $\Lambda$.
That is, each quantum state $\rho_s$ is represented as a probability distribution $P(\lambda |s)$ over the ontic states, $\lambda \in \Lambda$, of the system, and each measurement outcome, associated with quantum effect $E_{y|t}$, is represented by a conditional probability distribution $P(y|t\lambda)$ describing the probability of outcome $y$ occuring given that the measurement was $t$ and that the ontic state was $\lambda$.

An ontological model respects the assumption of generalized noncontextuality if the ontological representations of procedures respect the operational equivalences that hold among these. 
For the operational equivalences of Eq.~\eqref{opequivs}, generalized noncontextuality implies that
\begin{align}\label{ncoriginal}
&\sum_s \alpha_s^{(a)} P(\lambda|s) = 0\,, \quad
&\sum_{y,t} \beta_{yt}^{(b)} P(y|t,\lambda)=0 \,,
\end{align}
for all $\lambda \in\Lambda$,
  $a\in O_P$ and $b\in O_M$.

The correlations $P(y|s,t)$ that can be realized as $P(y|s,t) =\sum_{\lambda} P(y|t,\lambda) P(\lambda|s)$  for $P(y|t,\lambda)$ and $P(\lambda|s)$  satisfying Eq.~\eqref{ncoriginal}, 
 i.e., those that are {\em noncontextually realizable} for the given operational equivalences, form a polytope.  The facets of this polytope are examples of {\em noncontextuality inequalities}~\cite{Schmid2018}.
 One obtains a proof of contextuality whenever the observed quantum correlations $P^{\rm (q)}(y|s,t)$ violate at least one of these facet-defining noncontextuality inequalities. In such cases, there is no noncontextual ontological model that can reproduce the quantum correlations.
 
 There are many known examples of such proofs~\cite{POM,operationalks,Mazurek2016,Kunjwal16,kunjwal2018from, Kunjwal19,Kunjwal20,RAC,RAC2,Saha_2019,comp1,comp2,schmid2021only,schmid2018contextual,cloningcontext,contextmetrology, AWV, KLP19, YK20}; however, to the best of our knowledge, all previous proofs have involved a set of measurements that manifest some incompatibility.

{\bf Measurement incompatibility---}
Compatibility for generic quantum measurements---both sharp and unsharp---is 
defined in terms of joint simulability~\cite{busch1997operational}.
For the case of discrete outcome spaces, it can be expressed as follows~\cite{parable,guerini2017operational}: 
the measurements associated to a set of POVMs $\{\{E_{y|t}\}_y\}_t$ are said to be \emph{compatible} if there exists a single measurement, described by a POVM $\{G_z\}_z$, and a stochastic post-processing $P(y|t,z)$ such that
\beq
E_{y|t} = \sum_z P(y|t,z) G_z
\eeq
for all $y,t$. In this case, each measurement $\{E_{y|t}\}_y$ can be simulated by first measuring $\{G_z\}_z$ and then implementing a post-processing of the outcome statistics by $P(y|t,z)$.\footnote{ An equivalent definition of compatibility can be obtained by replacing stochastic post-processing by coarse-graining (see, e.g., Lemma 1 of Ref.~\cite{guerini2017operational}).  For sharp measurements, it is well known that this definition of compatibility reduces to commutation of the associated Hermitian operators (see Proposition 8 of Ref.~\cite{heinosaari2008notes} and Theorem 1 of Ref.~\cite{Kunjwal14} for proofs of some generalizations of this result).}
If such a measurement and post-processing do not exist, then the set of measurements is said to be \emph{incompatible}.

Note that a set consisting of a single measurement is trivially compatible.

{\bf Generalized noncontextuality no-go theorems without incompatibility---}
We now construct a class of operational scenarios that allow a proof of contextuality without making use of any measurement incompatibility.

We do so by starting with a prepare-measure scenario of the type described above---with a set of preparations associated to states $\{ \rho_s\}_s$,  a set of measurements associated to POVMs $\{ \{ E_{y|t}\}_y\}_t$, and operational equivalences described by Eq.~\eqref{opequivs}---and we implement a modification on the measurement side. 
Specifically, the modified scenario involves a single measurement which is obtained from the set of measurements in the original scenario by a procedure we term `flag-convexification'. 
 One flag-convexifies  the set of measurements 
  by randomly sampling a setting $t \in T$ according to some probability distribution $P(t)$, then implementing the measurement for that setting, $\{E_{y|t}\}_y$, and finally outputting both $y$ and $t$, so that $(y,t)\in Y\times T$ constitutes the outcome of the new effective measurement.  The terminology for the procedure stems from the fact that one is taking a convex mixture of the measurements in the original scenario, but one wherein the choice of measurement is not forgotten but rather flagged, i.e., copied and fed forward to be included in the output. 

To make the argument as simple as possible, we consider uniform sampling, i.e., $P(t)=\frac{1}{|T|}$, where $|T|$ is the cardinality of the set $T$ of possible settings.  (In our companion paper~\cite{companionincompatibility}, we show that any distribution with full support also works.) This simple version of flag-convexification is depicted in Fig.~\ref{fig1}.  
The single measurement in the flag-convexified scenario is associated to a POVM $\{\widetilde{E}_{y,t}\}_{y,t}$ defined by
\begin{equation} \label{effectranslation}
    \widetilde{E}_{y,t} := \frac{1}{|T|} E_{y|t}.
\end{equation} 
It is straightforward to check that this is a valid POVM.
For ease of bookkeeping, effects and conditional probability distributions which refer to the flag-convexified scenario will be denoted with a tilde.

\begin{figure}[htb!] 
\centering
\begin{equation*}
\begin{tikzpicture}
	\begin{pgfonlayer}{nodelayer}
		\node [style=none] (0) at (-0.5, -1) {};
		\node [style=none] (1) at (-0.5, 1) {};
		\node [style=none] (2) at (-1, -1) {};
		\node [style=none] (3) at (-1, -2) {};
		\node [style=none] (4) at (1, -2) {};
		\node [style=none] (5) at (1, -1) {};
		\node [style=none] (7) at (-1, 2) {};
		\node [style=none] (8) at (-1, 1) {};
		\node [style=none] (9) at (1, 1) {};
		\node [style=none] (10) at (1, 2) {};
		\node [style=none] (13) at (0.5, 2) {};
		\node [style=none] (14) at (0.5, 2.5) {};
		\node [style=none] (15) at (0.5, 0.5) {};
		\node [style=none] (16) at (0.5, 1) {};
		\node [style=none] (19) at (0.5, -2.5) {};
		\node [style=none] (20) at (0.5, -2) {};
		\node [style=right label] (21) at (0.5, -2.25) {$S$};
		\node [style=right label] (23) at (0.5, 2.25) {$Y$};
		\node [style=right label] (24) at (0.5, 0.75) {$T$};
		\node [style=right label] (25) at (-0.5, 0) {$\mathcal{H}$};
		\node [style=none] (26) at (0, -1.5) {$\{\rho_s\}$};
		\node [style=none] (27) at (0, 1.5) {$\{E_{y|t}\}$};
		\node [style=left label] (28) at (-1, -3) {$(a)$};
	\end{pgfonlayer}
	\begin{pgfonlayer}{edgelayer}
		\draw [qWire] (1.center) to (0.center);
		\draw (2.center) to (5.center);
		\draw (5.center) to (4.center);
		\draw (4.center) to (3.center);
		\draw (3.center) to (2.center);
		\draw (7.center) to (10.center);
		\draw (10.center) to (9.center);
		\draw (9.center) to (8.center);
		\draw (8.center) to (7.center);
		\draw [cWire] (14.center) to (13.center);
		\draw [cWire] (16.center) to (15.center);
		\draw [cWire] (20.center) to (19.center);
	\end{pgfonlayer}
\end{tikzpicture}}\ \quad \stackrel{\rm flag-conv.}{\longrightarrow}\ \quad  %
\begin{tikzpicture}
	\begin{pgfonlayer}{nodelayer}
		\node [style=none] (0) at (-0.5, -1) {};
		\node [style=none] (1) at (-0.5, 1) {};
		\node [style=none] (2) at (-1, -1) {};
		\node [style=none] (3) at (-1, -2) {};
		\node [style=none] (4) at (1, -2) {};
		\node [style=none] (5) at (1, -1) {};
		\node [style=none] (7) at (-1, 2) {};
		\node [style=none] (8) at (-1, 1) {};
		\node [style=none] (9) at (1, 1) {};
		\node [style=none] (10) at (1, 2) {};
		\node [style=none] (13) at (0, 2) {};
		\node [style=none] (14) at (0, 2.5) {};
		\node [style=none] (19) at (0.5, -2.5) {};
		\node [style=none] (20) at (0.5, -2) {};
		\node [style=right label] (21) at (0.5, -2.25) {$S$};
		\node [style=right label] (23) at (0, 2.25) {$Y$};
		\node [style=right label] (25) at (-0.5, 0) {$\mathcal{H}$};
		\node [style=none] (26) at (0, -1.5) {$\{\rho_s\}$};
		\node [style=none] (27) at (0, 1.5) {$\{\widetilde{E}_{y,t}\}$};
		\node [style=none] (31) at (0.75, 2) {};
		\node [style=none] (32) at (0.75, 2.5) {};
		\node [style=right label] (33) at (0.75, 2.25) {$T$};
		\node [style=left label] (34) at (-1, -3) {$(b)$};
	\end{pgfonlayer}
	\begin{pgfonlayer}{edgelayer}
		\draw [qWire] (1.center) to (0.center);
		\draw (2.center) to (5.center);
		\draw (5.center) to (4.center);
		\draw (4.center) to (3.center);
		\draw (3.center) to (2.center);
		\draw (7.center) to (10.center);
		\draw (10.center) to (9.center);
		\draw (9.center) to (8.center);
		\draw (8.center) to (7.center);
		\draw [cWire] (14.center) to (13.center);
		\draw [cWire] (20.center) to (19.center);
		\draw [cWire] (32.center) to (31.center);
	\end{pgfonlayer}
\end{tikzpicture}}\ \ :=\ \ %
\begin{tikzpicture}
	\begin{pgfonlayer}{nodelayer}
		\node [style=none] (0) at (-0.5, -1.75) {};
		\node [style=none] (1) at (-0.5, 1) {};
		\node [style=none] (2) at (-1, -1.75) {};
		\node [style=none] (3) at (-1, -2.75) {};
		\node [style=none] (4) at (1, -2.75) {};
		\node [style=none] (5) at (1, -1.75) {};
		\node [style=none] (7) at (-1, 2) {};
		\node [style=none] (8) at (-1, 1) {};
		\node [style=none] (9) at (1, 1) {};
		\node [style=none] (10) at (1, 2) {};
		\node [style=none] (13) at (0.5, 2) {};
		\node [style=none] (14) at (0.5, 2.5) {};
		\node [style=none] (15) at (0.5, 1) {};
		\node [style=none] (16) at (0.5, 1) {};
		\node [style=none] (19) at (0.5, -3.25) {};
		\node [style=none] (20) at (0.5, -2.75) {};
		\node [style=right label] (21) at (0.5, -3) {$S$};
		\node [style=right label] (23) at (0.5, 2.25) {$Y$};
		\node [style=right label] (25) at (-0.5, 0) {$\mathcal{H}$};
		\node [style=none] (26) at (0, -2.25) {$\{\rho_s\}$};
		\node [style=none] (27) at (0, 1.5) {$\{E_{y|t}\}$};
		\node [style=none] (31) at (1.5, 1) {};
		\node [style=none] (32) at (1.5, 2.5) {};
		\node [style=right label] (33) at (1.5, 2.25) {$T$};
		\node [style=white dot] (34) at (1, 0.5) {};
		\node [style=none] (35) at (1, -0.25) {};
		\node [style=right label] (36) at (1, 0) {$T$};
		\node [style=none] (37) at (1, -0.675) {\scriptsize$\frac{1}{|T|}$};
		\node [style=none] (38) at (0.5, -0.25) {};
		\node [style=none] (39) at (1.5, -0.25) {};
	\end{pgfonlayer}
	\begin{pgfonlayer}{edgelayer}
		\draw [qWire] (1.center) to (0.center);
		\draw (2.center) to (5.center);
		\draw (5.center) to (4.center);
		\draw (4.center) to (3.center);
		\draw (3.center) to (2.center);
		\draw (7.center) to (10.center);
		\draw (10.center) to (9.center);
		\draw (9.center) to (8.center);
		\draw (8.center) to (7.center);
		\draw [cWire] (14.center) to (13.center);
		\draw [cWire] (16.center) to (15.center);
		\draw [cWire] (20.center) to (19.center);
		\draw [cWire] (32.center) to (31.center);
		\draw [cWire, in=150, out=-90] (15.center) to (34);
		\draw [cWire] (34) to (35.center);
		\draw [cWire, in=-90, out=30] (34) to (31.center);
		\draw (38.center) to (39.center);
		\draw [bend left=90, looseness=3.75] (39.center) to (38.center);
	\end{pgfonlayer}
\end{tikzpicture}} 
\end{equation*}
\caption{(a) The original prepare-measure scenario. (b) The flag-convexified scenario, where the white dot represents the copying of $T$.} \label{fig1}
\end{figure}

Note that the correlations in the original scenario are naturally associated with a conditional probability distribution of the form $P^{\rm (q)}(y|s,t):=\mathsf{tr}(E_{y|t}\rho_s)$ where $t$ is on the right-hand-side of the conditional, while the correlations in the flag-convexified scenario are naturally associated with a conditional probability distribution of the form $\widetilde{P}^{\rm (q)}(y,t|s):=\mathsf{tr}(\widetilde{E}_{y,t} \rho_s)$ where $t$ is on the left-hand-side of the conditional. 
Using Eq.~\eqref{effectranslation} and the linearity of the Born rule, 
these are related simply by
\begin{align}\label{pxpyp}
	\widetilde{P}^{\rm(q)}(y,t|s)
	= \frac{1}{|T|}P^{\rm(q)}(y|s,t).
\end{align}

Since the set of preparations in the flag-convexified scenario is identical to the set in the original scenario, it follows that the operational equivalences for the preparations are unchanged relative to Eq.~\eqref{opequivs},
\begin{align}\label{OEP}
\sum_s \alpha_s^{(a)} \rho_s = 0,
\end{align}
 for all $a\in O_P$. 
Consider now the operational equivalences among the effects in the new scenario. Substituting Eq.~\eqref{effectranslation}
 into Eq.~\eqref{opequivs}, one finds that 
 the operational equivalences in the flag-convexified scenario are given by 
\begin{align}\label{OEM}
&\sum_{y,t} \beta_{yt}^{(b)} \widetilde{E}_{y,t}=0
\end{align}
 for all  $b\in O_M$. 
So the effects $\widetilde{E}_{y,t}$ in the flag-convexified scenario satisfy linear constraints of exactly the same form as those satisfied by the effects $E_{y|t}$ from the original scenario.

An ontological representation of the flag-convexified scenario represents each state $\rho_s$ by some probability  distribution $\widetilde{P}(\lambda|s)$ and each effect $\widetilde{E}_{y,t}$ by some conditional probability distribution $ \widetilde{P}(y,t|\lambda)$.
Given the operational equivalences of Eqs.~\eqref{OEP} and \eqref{OEM}, the assumption of 
 noncontextuality implies that $\widetilde{P}(\lambda|s)$ and $ \widetilde{P}(y,t|\lambda)$ must satisfy
\begin{align} \label{ncfc}
&\sum_s \alpha_s^{(a)} \widetilde{P}(\lambda|s) = 0, \
&\sum_{y,t} \beta_{yt}^{(b)} \widetilde{P}(y,t|\lambda)=0,
\end{align}
for all $\lambda \in \Lambda$, $a\in O_P$ and $b\in {O}_M$,  which are seen to be of the same form as Eq.~\eqref{ncoriginal}.

In the appendix, we prove that there exists a noncontextual ontological model for the original scenario if and only if there exists a noncontextual ontological model for the flag-convexified scenario. 
This fact implies that every no-go theorem for noncontextuality in a prepare-measure scenario involving incompatibility (of which there are many) can be transformed via flag-convexification into a no-go theorem in a scenario that involves only a single measurement (and hence involves no incompatibility). 
This establishes Theorem~\ref{thm1}.

 {\bf An example not built from processings of incompatible sharp measurements---}
The opportunity for proving generalized contextuality without incompatibility arises from the fact that there can be nontrivial operational equivalences between the outcomes of one and the same measurement if the latter is unsharp. It is natural to ask whether this opportunity {\em only} presents itself when the unsharp measurement is a flag-convexification (or stochastic post-processing) of a set of incompatible sharp measurements and when the operational equivalences in the former are implied by the conventional variety of operational equivalences in the latter.  If it did, then one might dismiss proofs {\em without} incompatibility as simply proofs {\em with} incompatibility ``in disguise''.  As we now demonstrate, the question is answered in the negative---there are genuinely novel types of proofs of contextuality for unsharp measurements.

Consider a prepare-measure scenario 
with five
preparations,  associated with the 
set of normalized rank-1 projectors $\{ \rho_s = |\psi_s \rangle \langle \psi_s| \}_{s=0}^4$ where
\begin{align}
|\psi_s\rangle := \cos\left(\frac{\pi}{5}s \right) |0\rangle + \sin\left(\frac{\pi}{5}s \right) |1\rangle 
\end{align}
and a single 5-outcome measurement, associated with the POVM $\{ E_y \}_{y=0}^4$ where the effects are defined as ${E_y \coloneqq \tfrac{2}{5} \rho_y}$ (i.e., subnormalized versions of the projectors onto the states) and thus sum to the identity.

These quantum states and effects can be represented as real-valued vectors in the space of Hermitian operators using the basis of Pauli operators.  Because they all have zero component of the $Y$ Pauli, they can be represented in the three-dimensional space spanned by $I$, $Z$, and $X$ Paulis.  We provide these representations in Figs.~\ref{thenewfig}(a) and (b) respectively.  For five vectors in a three-dimensional space,  there are necessarily linear dependences amongst them, which can be captured by a pair of  
equations.
 For the five states, these are interpreted as a pair of operational equivalences. They can be expressed as:
\begin{align}\label{eq:oe1}
\begin{split}
 & \rho_0 - q \rho_1 + q \rho_2  - \rho_3 =0 \,, \\
&\rho_1 - q \rho_2 + q \rho_3  - \rho_4=0\,, \\
\end{split}		
\end{align} 
where $q$ is the golden ratio, namely $q = 2 \cos \left({\pi/5}\right)=(1+\sqrt{5})/2$.
The five effects satisfy the same operational equivalence relations (substituting $E_i$ for $\rho_i$ in Eq.~\eqref{eq:oe1}) because they are equal to the states up to a normalization factor.

\begin{figure}
\begin{center}
\subfigure[State preparations]{
\begin{tikzpicture}[scale=3]

    \draw [->, color=black,dotted] (0,-0.5 ) -- (0,1.7) node (iaxis) [above] {$\hat{e}_\mathrm{I}$};
    \draw [->, color=black] (190:1) -- (10:1) node (zaxis) [right] {$\hat{e}_\mathrm{Z}$};
    \draw [->, color=black] (160:1) -- (-20:1) node (xaxis) [right] {$\hat{e}_\mathrm{X}$};
    \draw[color=black] (0,-0.5) -- (0,0);
    \draw[color=black] (0,0.7) -- (0,1.7);

\draw [rotate=0, x radius=1cm, y radius=0.3cm, delta angle=360, color=gray] (1,0.7)
    arc [start angle=0]
    node (b1) [pos=0.0,draw,scale=0.1,shape=circle] {}
    node (b2) [pos=0.5,draw,scale=0.1,shape=circle] {}
    node (e1) [pos=0.50 ,draw=panblue,fill, color=panblue, scale=.2, shape=circle] {}
    node (e2) [pos=0.70,draw=panblue,fill, color=panblue, scale=.2, shape=circle] {}
    node (e3) [pos=0.90,draw=panblue,fill, color=panblue, scale=.2, shape=circle] {}
    node (e4) [pos=0.10,draw=panblue,fill, color=panblue, scale=.2, shape=circle] {}
	node (e5) [pos=0.30,draw=panblue,fill, color=panblue, scale=.2, shape=circle] {};
    	
\node at ($(0,0)!1.17!(e4)$) {\color{panblue}{$\ket{0}\hskip -0.07cm \bra{0}$}};    	
    	
    \draw[->,thick,color=panblue] (0,0) -- (e1);    
    \draw[->,thick,color=panblue] (0,0) -- (e2); 
    \draw[->,thick,color=panblue] (0,0) -- (e3); 
    \draw[->,thick,color=panblue] (0,0) -- (e4); 
    \draw[->,thick,color=panblue] (0,0) -- (e5); 
    
    \draw[color=carmine] (e1)--(e2)--(e3)--(e4)--(e5)--(e1);


\path[rotate=0, x radius=1cm, y radius=0.3cm, delta angle=360, name path=p1] (1,0.7)   arc [start angle=0];
\node (c1) at ($(b1)!0.15!(0,0)$) {};
\path[name path=p2] (0,0) -- (c1);
\node (c2) at ($(b2)!0.15!(0,0)$) {};
\path[name path=p3] (0,0) -- (c2);

\path [name intersections={of=p1 and p2}] ;
\draw[gray!50!white] (0,0) -- (intersection-1);
\draw[gray!50!white,dotted,thick] (b1) -- (intersection-1);

\path [name intersections={of=p1 and p3}] ;
\draw[gray!50!white] (0,0) -- (intersection-1);
\draw[gray!50!white,dotted,thick] (b2) -- (intersection-1);

\end{tikzpicture}}
\hspace{0.5cm}
\subfigure[The single measurement]{
\begin{tikzpicture}[scale=3]
    \draw [->, color=black,dotted] (0,-0.5 ) -- (0,1.7) node (iaxis) [above] {$\hat{e}_\mathrm{I}$};
    \draw [->, color=black] (190:1) -- (10:1) node (zaxis) [right] {$\hat{e}_\mathrm{Z}$};
    \draw [->, color=black] (160:1) -- (-20:1) node (xaxis) [right] {$\hat{e}_\mathrm{X}$};
    \draw[color=black] (0,-0.5) -- (0,0);
    \draw[color=black] (0,1.5) -- (0,1.7);

\draw [rotate=0, x radius=1cm, y radius=0.3cm, delta angle=360, color=gray,dotted] (1,0.7)
    arc [start angle=0]
    node (b1) [pos=0.0,draw=gray,fill,color=gray,,scale=0.1,shape=circle] {}
    node (b2) [pos=0.5,draw=gray,fill,color=gray,,scale=0.1,shape=circle] {}
    node (e1) [pos=0.50 ,draw=gray,fill,color=gray,scale=.2, shape=circle] {}
    node (e2) [pos=0.70,draw=gray,fill,color=gray,scale=.2, shape=circle] {}
    node (e3) [pos=0.90,draw=gray,fill,color=gray,scale=.2, shape=circle] {}
    node (e4) [pos=0.10,draw=gray,fill,color=gray,scale=.2, shape=circle] {}
	node (e5) [pos=0.30,draw=gray,fill,color=gray,scale=.2, shape=circle] {};

\node at ($(0,0)!1.17!(e4)$) {\color{panblue}{$\ket{0}\hskip -0.07cm \bra{0}$}};

\path[rotate=0, x radius=1cm, y radius=0.3cm, delta angle=360, name path=p1] (1,0.7)   arc [start angle=0];
\node (c1) at ($(b1)!0.15!(0,0)$) {};
\path[name path=p2] (0,0) -- (c1);
\node (c2) at ($(b2)!0.15!(0,0)$) {};
\path[name path=p3] (0,0) -- (c2);

\path [name intersections={of=p1 and p2}] ;
\draw[gray!50!white] (0,0) -- (intersection-1);
\draw[gray!50!white,dotted,thick] (b1) -- (intersection-1);

\path [name intersections={of=p1 and p3}] ;
\draw[gray!50!white] (0,0) -- (intersection-1);
\draw[gray!50!white,dotted,thick] (b2) -- (intersection-1);
 
\draw[gray!50!white] (0,1.5) -- (b2);
\draw[gray!50!white] (0,1.5) -- (b1);

\draw [rotate=0, x radius=1cm, y radius=0.3cm, delta angle=180, color=gray] (-1,0.7) arc [start angle=180];


\node[draw=panblue,fill, color=panblue, scale=.2, shape=circle] (ep4) at ($(e4)!0.6!(0,0)$) {};	
\node[draw=panblue,fill, color=panblue, scale=.2, shape=circle] (ep5) at ($(e5)!0.6!(0,0)$) {};	
\node[draw=panblue,fill, color=panblue, scale=.2, shape=circle] (ep1) at ($(e1)!0.6!(0,0)$) {};	
\node[draw=panblue,fill, color=panblue, scale=.2, shape=circle] (ep2) at ($(e2)!0.6!(0,0)$) {};	
\node[draw=panblue,fill, color=panblue, scale=.2, shape=circle] (ep3) at ($(e3)!0.6!(0,0)$) {};	

	\draw[->,thick,color=panblue] (0,0) -- (ep1);    
    \draw[->,thick,color=panblue] (0,0) -- (ep2); 
    \draw[->,thick,color=panblue] (0,0) -- (ep3); 
    \draw[->,thick,color=panblue] (0,0) -- (ep4); 
    \draw[->,thick,color=panblue] (0,0) -- (ep5); 
    
    \draw[thick,dotted,color=panblue] (e1) -- (ep1);    
    \draw[thick,dotted,color=panblue] (e2) -- (ep2); 
    \draw[thick,dotted,color=panblue] (e3) -- (ep3); 
    \draw[thick,dotted,color=panblue] (e4) -- (ep4); 
    \draw[thick,dotted,color=panblue] (e5) -- (ep5); 
    
    \draw[color=carmine] (ep1)--(ep2)--(ep3)--(ep4)--(ep5)--(ep1);

\end{tikzpicture}}
\end{center}
\caption{\textbf{An explicit example of contextuality without incompatibility.}
Bloch-vector representation of (a) the five quantum state preparations, and (b) the five outcomes of the single unsharp measurement.
Since all operators in the example involve no component of the Pauli-$Y$ matrix, we depict all operators by their 3-dimensional vector representation $(a_I,a_X,a_Z)$ defined by $A = a_I \, I + a_X \, X + a_Z \, Z$, where $I$, $X$, and $Z$ are the other three Pauli operators.}
\label{thenewfig}
\end{figure}

By applying the linear programming techniques of Ref.~\cite{Krishna_2017} to this scenario, we can compute 
facet-defining noncontextuality inequalities. 
(Further details are provided in the appendix.)

Consider a conditional probability distribution $P(y|s)$ (whose elements we will abbreviate as $p_{y|s}$) and the following linear combination of these elements:
\begin{align}\label{NCinequality}
\mathcal{C}:= q \left(p_{1|0}+ p_{1|2}\right)+(q-1)p_{2|0}+p_{0|2}-(q+1)p_{1|1}\,.
\end{align}
An inequality that defines a facet of the polytope of noncontextually realizable conditionals $P(y|s)$ is found to be:
\begin{align}\label{keyncineq2}
\mathcal{C} \geq 0\,.
\end{align} 

Meanwhile, the quantumly realizable correlations for the scenario described above yield
\begin{align}\label{QviolationNCineq}
\mathcal{C}^{\rm (q)} = \frac{q^2 -4}{10}
\approx -0.138\,,
\end{align}
which violates the noncontextuality inequality of Eq.~\eqref{keyncineq2}, thereby proving that the quantum scenario does not admit of a noncontextual model. 

Finally, we demonstrate that the 5-outcome POVM used in this proof cannot be understood as the flag-convexification of a set of  projector-valued measures (PVMs), nor even as a stochastic  processing of a set of PVMs. (Recall that sharp measurements in quantum theory are represented by PVMs.)
Suppose that the 5-outcome POVM $\{E_0, \dots, E_4\}$ {\em could} be obtained by stochastic processing of a set $\{\mathcal{P}^{(\alpha)}\}_{\alpha}$ of distinct binary-outcome qubit PVMs, $\mathcal{P}^{(\alpha)} := \{ \Pi^{(\alpha)}_0,\Pi^{(\alpha)}_1\}$.
Specifically, this would mean that $E_y = \sum_{j,k} P(y|j,k) \left( \sum_{\alpha} \Pi_j^{(\alpha)} P(\alpha,k) \right)$
 for some auxiliary variable $k$, distribution $P(\alpha, k)$ and conditional $P(y|j,k)$.
  However, because each effect $E_y$ is itself rank-1 and because all the $\Pi^{(\alpha)}_j$ are distinct, these 
 sums (which are non-negative by virtue of the weights $P(y|j,k)P(\alpha,k)$ being non-negative)
   can each contain only a single rank-1 projector.  Moreover, because post-selection is not allowed in such a processing, each projector must appear in the positive sum yielding {\em some} effect. It follows that there  must be a one-to-one association between the five effects and the full set of projectors.  But this yields a contradiction. One way to see this is that no two effects are orthogonal, while the complementary rank-1 projectors appearing in a single PVM,  $\Pi^{(\alpha)}_0$ and $\Pi^{(\alpha)}_1$,
   {\em are} orthogonal. (Alternatively, one can note simply that there are an odd number of effects in the POVM and an even number of projectors in the set of PVMs.)

\textbf{Related work---} Ref.~\cite{singh2021revealing} has also noted that there is an opportunity for leveraging operational equivalences among the elements of a single unsharp measurement in proofs of 
generalized contextuality\footnote{In fact, Ref.~\cite{singh2021revealing} claims that the notion of noncontextuality proposed in Ref.~\cite{gencontext} must be supplemented with an {\em additional} constraint: that if two effects $E$ and $E'$ satisfy $E' =sE$ where $s\in [0,1]$,  then the conditional probability distributions that represent them, $P(E|\lambda)$ and $P(E'|\lambda)$, must satisfy $P(E'|\lambda)= s P(E|\lambda)$.   However, this constraint is not an innovation to the notion of noncontextuality, but rather is known to follow from the representation of post-processing within any ontological model, as noted in Ref.~\cite{determinism} (see the proof of condition NC3 therein).}---indeed, the construction they consider is 
an instance of  what we here termed flag-convexification. 
 The analysis in Ref.~\cite{singh2021revealing},  however, makes use of an auxiliary assumption, namely, that the probability distribution over ontic states after a measurement is proportional to the response function of the effect that was observed.
This assumption was not shown to follow from noncontextuality, so Ref.~\cite{singh2021revealing} does not establish our result.\footnote{ Additionally, we were not able to show that this assumption follows from noncontextuality. Furthermore,  if one reconceptualizes their scenario as a prepare-measure experiment, so that one can apply the techniques of Ref.~\cite{Schmid2018}, we do not obtain
the inequalities from Ref.~\cite{singh2021revealing} as valid noncontextuality inequalities.}

 It is also worth drawing the parallel between our work and that of Fritz~\cite{fritz2012beyond,fritz2016beyond}, which established that there are Bell-like 
causal compatibility inequalities~\cite{fritz2012beyond,wood2015lesson,Wolfe2016inflation,tavakoli2021bell} admitting of quantum violations even in scenarios where each party implements only a single measurement, implying that incompatibility is also not necessary for exhibiting quantumness in such cases.

{\bf Conclusions---}
We have demonstrated that incompatibility is not required for proofs of generalized contextuality.
 This fact shows a new sense in which the failure of generalized noncontextuality can be proven in a broader range of scenarios than the failure of Kochen-Specker noncontextuality.  This result also opens the door to the possibility of finding new quantum advantages for information processing by considering operational scenarios with no incompatibility, some of which might have previously been overlooked by virtue of mistakenly being thought to offer no opportunity for nonclassical phenomena. 

Our companion paper, Ref.~\cite{companionincompatibility}, generalizes the results in this paper in a number of ways. First, the arguments therein  are formulated in the context of  arbitrary generalized probabilistic theories rather than just quantum theory.  This means that if a given generalized probabilistic theory admits of a proof of contextuality in a prepare-measure scenario, then one can find such a proof involving only a single measurement.
Second, we demonstrate the possibility of proofs of contextuality (for both quantum theory and GPTs) in scenarios that have only a single preparation device as well as a single measurement device.
Such scenarios have no external inputs at all. Hence, in addition to not requiring any incompatibility, they also do not require the freedom of choice assumption---roughly,  that one can choose one's external inputs freely. 
Finally, we show that if one's detectors are inefficient, in that they have probability $p$ of performing ideally and probability $1-p$ of failing to give any outcome, then one can still witness contextuality, even for {\em arbitrary} inefficiencies,  i.e., any $p>0$.  That is, there is no detector loophole for tests of generalised noncontextuality.

\textbf{Acknowledgements.---} This research was supported by Perimeter Institute for Theoretical Physics. Research at Perimeter Institute is supported by the Government of Canada through the Department of Innovation, Science and Economic Development Canada and by the Province of Ontario through the Ministry of Research, Innovation and Science. 
DS, and ABS acknowledge support by the Foundation for Polish Science (IRAP project, ICTQT, contract no.2018/MAB/5, co-financed by EU within Smart Growth Operational Programme). JHS was supported by the
National Science Centre, Poland (Opus project, Categorical
Foundations of the Non-Classicality of Nature, project
no. 2021/41/B/ST2/03149). RK is supported by the Charg{\'e} de
Recherche fellowship of the Fonds de la Recherche Scientifique
- FNRS (F.R.S.-FNRS), Belgium.
All of the diagrams within this manuscript were prepared using TikZit. Figures were prepared using Tikz.

\bibliographystyle{apsrev4-2-wolfe}
\setlength{\bibsep}{3pt plus 3pt minus 2pt}
\nocite{apsrev42Control}
\bibliography{bib}

\appendix

\section{Proof of Theorem 1}

We prove Theorem 1 by demonstrating that the flag-convexified scenario (introduced in the main text) admits of a non-contextual ontological model iff the original scenario does so. Specifically, this requires proving that there exist $P(\lambda|s)$ and $P(y|t,\lambda)$ satisfying the constraints of Eq.~(3) from the main text and 
which reproduce the quantum statistics in the original scenario, i.e., 
\beq\label{cprthry}
P^{\rm (q)}(y|s,t) = \sum_{\lambda}P(y|t,\lambda) P(\lambda|s)
\eeq
  {\em if and only if } there exist $\widetilde{P}(\lambda|s)$ and $\widetilde{P}(y,t|\lambda)$ satisfying the constraints of Eq.~(9) from the main text and reproducing the quantum statistics in the flag-convexified scenario, i.e., 
  \beq\label{pxyst2}
  \widetilde{P}^{\rm (q)}(y,t|s) = \sum_{\lambda} \widetilde{P}(y,t|\lambda) \widetilde{P}(\lambda|s).
  \eeq
It then straightforwardly follows that every no-go theorem for noncontextuality in a prepare-measure scenario involving incompatibility can be transformed (via flag-convexification) into a no-go theorem involving a single measurement, that is, into a no-go theorem without incompatibility.

We begin with the forward implication.  It suffices to take
\begin{align}\label{jjj1}
&\widetilde{P}(\lambda|s):=P(\lambda|s), \
&\widetilde{P}(y,t|\lambda):=\frac{1}{|T|}P(y|t,\lambda).
\end{align}
It is straightforward to check that $\widetilde{P}(y,t|\lambda)$ is a valid conditional probability distribution.
From the fact that $P(\lambda|s)$ and $P(y|t,\lambda)$ satisfy the noncontextuality constraints in Eq.~(3) from the main text, it follows immediately that $\widetilde{P}(\lambda|s)$ and $\widetilde{P}(y,t|\lambda)$ satisfy the noncontextuality constraints in Eq.~(9) from the main text.   
 Eq.~\eqref{pxyst2} then follows from Eq.~\eqref{cprthry} by making use of Eqs.~(6) from the main text and \eqref{jjj1}, i.e.,
\begin{align}
\widetilde{P}^{\rm(q)}(y,t|s)   &= \frac{1}{|T|}P^{\rm (q)}(y|s,t)\\
&=\frac{1}{|T|} \sum_{\lambda}  P(y|t,\lambda)P(\lambda|s)\\
&=\sum_{\lambda} \widetilde{P}(y|t,\lambda) \widetilde{P}(\lambda|s).
\end{align}

The reverse implication follows by an analogous argument, but where one defines $P(\lambda|s) := \widetilde{P}(\lambda|s)$ and $P(y|t,\lambda):=|T| \widetilde{P}(y,t|\lambda)$.

\section{Proof of noncontextuality inequality}

We seek to put bounds on the correlations in the prepare-measure experiment.  These are given in the text in terms of $P(y|s)$, where $y$ runs over the five outcomes of the measurement and $s$ runs over the five preparations.  In an ontological model, we have
\begin{align}
P(y|s)= \sum_{\lambda} P(y|\lambda) P(\lambda|s).
\end{align}
But by implementing a Bayesian inversion between $\lambda$ and $s$, this can also be expressed as
\begin{align}\label{eq:gen}
P(y|s)= \sum_{\lambda} P(y|\lambda) P(s|\lambda) P(\lambda) P(s)^{-1}.
\end{align}
We here assume a uniform prior over $s$, i.e., $P(s)=1/5$.  In this case, whatever linear dependences hold among the $P(\lambda|s)$ for a given $\lambda$, the same linear dependences hold among the $P(s|\lambda)$ for a given $\lambda$.  It will be convenient below to focus on the joint distribution $P(y,s)$, which takes the form
\begin{align}\label{eq:genjoint}
P(y,s)= \sum_{\lambda} P(y|\lambda) P(s|\lambda) P(\lambda).
\end{align}
For a uniform prior, this is related to $P(y|s)$ by a constant factor.

 We now derive the constraints on $P(y,s)$ that are implied by the assumption of noncontextuality and the operational equivalences that hold among the states and among the effects.

Recall that the assumption of noncontextuality means that operational equivalences of the form of Eq.~(2) in the main text imply corresponding constraints on the representations of states and effects in the ontological model, namely, Eqs.~(3) from the main text.  

The particular form of operational equivalences appearing in the pentagon example are
\begin{align}\label{eq:oestates2}
\begin{split}
 & \rho_0 - q \rho_1 + q \rho_2  - \rho_3 =0 \,, \\
&\rho_1 - q \rho_2 + q \rho_3  - \rho_4=0\,, \\
\end{split}
\end{align} 
where $q = 2 \cos \left({\pi/5}\right)=(1+\sqrt{5})/2$
(the golden ratio), and 
\begin{align}\label{eq:oeeffects2}
\begin{split}
 & E_0 -  q E_1 + q E_2   - E_3 =0 \,, \\
& E_1 - q E_2 + q E_3  - E_4=0\,. 
\end{split}
\end{align}

Given the five-fold symmetry of the set of states, the linear dependence relations described by the pair of equations in Eq.~\eqref{eq:oestates2} can be equivalently expressed by the image of this pair of equations under any cyclic permutation of the five states. Similarly, the five-fold symmetry of the set of effects implies that the linear dependence relations described by the pair of equations in Eq.~\eqref{eq:oeeffects2} are equivalent to those obtained by cyclic permutations of the effects.

Note that there is an intuitive justification of the operational equivalence in the first equation of Eq.~\eqref{eq:oestates2}: the ensemble consisting of pure states $\rho_0$ and $\rho_2$, drawn with probabilities $1/(q+1)$ and $q/(q+1)$ respectively, yields the same density operator as the ensemble consisting of pure states $\rho_1$ and $\rho_3$, drawn with probabilities $q/(q+1)$ and $1/(q+1)$, respectively:
\begin{align}\label{opequivpentagon}
\frac{1}{q+1}\rho_0 + \frac{q}{q+1}\rho_2 = \frac{q}{q+1}\rho_1 + \frac{1}{q+1}\rho_3.
\end{align}
This equality is easily verified by considering the geometry of the five states in the Bloch sphere (Fig.~\ref{convexmixtures}).  The weight of $\rho_0$ in the mixed state $\rho$ (indicated in Fig.~\ref{convexmixtures}) is proportional to the Euclidean distance between $\rho$ and $\rho_2$, while the weight of $\rho_2$ is proportional to the Euclidean distance between $\rho$ and $\rho_0$.  The ratio of these two Euclidean distances is $1/q$ where $q$ is the golden ratio, since this is one of the ways in which the golden ratio appears in the geometry of the pentagon.  It follows that the ratio of the weight of $\rho_0$ to the weight of $\rho_2$ in the mixed state $\rho$ is also $1/q$.  Requiring that these weights sum to unity then implies that they must be $1/(q+1)$ and $q/(q+1)$ respectively.  The fact that $\rho_3$ and $\rho_1$ respectively have the same Euclidean distances to $\rho$ as do $\rho_0$ and $\rho_2$  implies that $\rho_3$ and $\rho_1$ also appear with weights $1/(q+1)$ and $q/(q+1)$ in a convex decomposition of $\rho$.  Equating the two convex decompositions of $\rho$ yields Eq.~\eqref{opequivpentagon}.  

 The second operational equivalence relation in Eq.~\eqref{eq:oestates2} 
  is justified similarly, and the symmetry of these relations under cyclic permutations of the states is also evident from the geometry.  A similar analysis holds for the  operational equivalence relations in Eq.~\eqref{eq:oeeffects2}.

\begin{figure}[htbp]
\begin{center}

\begin{tikzpicture}[scale=3]

    \draw [->, color=gray] (180:1.3) -- (0:1.3) node (zaxis) [right] {\color{black}{$\hat{e}_\mathrm{X}$}};
    \draw [->, color=gray] (-90:1.3) -- (90:1.3) node (xaxis) [right] {\color{black}{$\hat{e}_\mathrm{Z}$}};

\draw [rotate=0, x radius=1cm, y radius=1cm, delta angle=360, color=gray] (1,0)
    arc [start angle=0]
    node (e1) [pos=0.25 ,draw=panblue,fill, color=panblue, scale=.2, shape=circle] {}
    node (e2) [pos=0.45, draw=panblue,fill, color=panblue, scale=.2, shape=circle] {}
    node (e3) [pos=0.65,draw=panblue,fill, color=panblue, scale=.2, shape=circle] {}
    node (e4) [pos=0.85,draw=panblue,fill, color=panblue, scale=.2, shape=circle] {}
	node (e5) [pos=0.05,draw=panblue,fill, color=panblue, scale=.2, shape=circle] {};

\draw[thick, color=panblue] (e1) -- (e2) -- (e3) -- (e4) -- (e5) -- (e1);

\node at (95:1.1) {$\rho_0$};
\node at (167:1.1) {$\rho_4$};
\node at (239:1.1) {$\rho_3$};
\node at (311:1.1) {$\rho_2$};
\node at (381:1.15) {$\rho_1$};

\draw[thick, color=carmine, name path=p1] (e1) -- (e4);
\draw[thick, color=carmine, name path=p2] (e5) -- (e3);

\path [name intersections={of=p1 and p2}] ;
\node[draw=carmine,fill, color=carmine, scale=.3, shape=circle] (rho1) at (intersection-1) {};

\node[ right = 0.5mm of rho1]  {\color{carmine}{$\rho$}};

\end{tikzpicture}

\caption{Understanding the operational equivalences holding among the states in terms of mixtures of states.}
\label{convexmixtures}
\end{center}
\end{figure}

 Generalized noncontextuality is the assumption that for each operational equivalence among states or among effects,
 one must posit a corresponding constraint on the ontological representation of these states and effects. 
Thus, in the concrete example considered here, for each $\lambda$, the $P(\lambda|s)$  satisfy linear dependence conditions parallel to those of Eq.~\eqref{eq:oestates2}.  Furthermore, as noted above, for each $\lambda$, the Bayesian inverses of these conditionals, $P(s|\lambda)$, must satisfy the same linear dependence conditions as the $P(\lambda|s)$, together with normalization.
Consequently, we have the constraints:
\begin{align}\label{eq:consstates}
\begin{split}
 & P(s=0|\lambda) - q P(s=1|\lambda) + q P(s=2|\lambda)   - P(s=3|\lambda) =0 \,, \\
& P(s=1|\lambda) - q P(s=2|\lambda) + q P(s=3|\lambda)  - P(s=4|\lambda)=0\,,\\
& \sum_s P(s|\lambda)=1.
\end{split}
\end{align} 

For the effects, generalized noncontextuality implies that the $P(y|\lambda)$ must satisfy linear dependence conditions parallel to those of Eq.~\eqref{eq:oeeffects2}, together with normalization, so that
\begin{align}\label{eq:conseffects}
\begin{split}
 &   P(y=0|\lambda) - q P(y=1|\lambda) + q P(y=2|\lambda)  - P(y=3|\lambda) =0 \,, \\
& P(y=1|\lambda) - q P(y=2|\lambda) + q P(y=3|\lambda) - P(y=4|\lambda)=0\,. \\
& \sum_y P(y|\lambda)=1.
\end{split}
\end{align} 

The constraints in Eq.~\eqref{eq:consstates} must hold for every ontic state $\lambda$.  If a probability distribution on $s$, i.e., $(P(s=0),P(s=1),P(s=2),P(s=3), P(s=4))$, satisfies these constraints, then it is termed a {\em noncontextual assignment} to $s$.  In a noncontextual model, every ontic state $\lambda$ makes such an assignment, which describes what one can retrodict about $s$ given $\lambda$.   

Similarly, if a probability distribution on $y$, i.e., $(P(y=0),P(y=1),P(y=2),P(y=3), P(y=4))$, satisfies the constraints in Eq.~\eqref{eq:conseffects}, then it is termed a {\em noncontextual assignment} to $y$.  Again, every ontic state $\lambda$ makes such an assignment in a noncontextual model, describing what one can {\em predict} about $y$ given $\lambda$.   

It is useful to consider the set of all possible noncontextual assignments to $s$.  By the linearity of the constraints, any convex mixture of a pair of noncontextual assignments is also a valid noncontextual assignment, so the set is convex.  In fact, it is a polytope contained within the 5-vertex simplex of all probability assignments over $s$.  Given that the possible noncontextual assignments to $y$ satisfy precisely the same constraints as the noncontextual assignments to $s$ (since Eqs.~\eqref{eq:consstates} and \eqref{eq:conseffects} have the same form), these describe precisely the same polytope.  We turn now to the description of this polytope.

Because of the invariance of the constraints under a cyclic permutation of the five components of the probability distribution,  it follows that any cyclic permutation of the components of a noncontextual assignment yields a noncontextual assignment.  

In fact, there are precisely five vertices of the polytope of noncontextual assignments, namely, the cyclic permutations of the following assignment:
\begin{align}\label{eq:NCassignment}
\frac{1}{\sqrt{5}}
(\; 1 \;,\; q^{-1} \;,\;0 \;,\;0 \;,\; q^{-1} \;).
\end{align}
  One can easily verify that this probability distribution 
 satisfies Eq.~\eqref{eq:consstates} (or Eq.~\eqref{eq:conseffects}) and satisfies normalization 
 (it suffices to recall that the golden ratio $q$ satisfies $q^{-1}=q-1$ and $1+2q^{-1}=\sqrt{5}$).  Extremality can be verified by mathematical software.\footnote{The presence of irrational coefficients complicates the task of identifying the extremal solutions to a set of linear constraints. For very small problems, such as the one here, however, it can be tackled by Mathematica\textsuperscript{\textregistered}.}

Let $\kappa$ vary over the five extremal noncontextual assignments to $y$
 and let $\kappa'$ vary over the five extremal noncontextual assignments to $s$.
For every ontic state $\lambda$, the noncontextual assignment to $y$
 must be a convex mixture of the extremal ones: $P(y|\lambda)=\sum_{\kappa}P(y|\kappa)P(\kappa|\lambda)$ for some $P(\kappa|\lambda)$. 
 Similarly, the noncontextual assignment to $s$
  by any ontic state $\lambda$ must be a convex mixture of the extremal ones: $P(s|\lambda)=\sum_{\kappa'} P(s|\kappa')P(\kappa'|\lambda)$ for some $P(\kappa'|\lambda)$.  It follows that Eq.~\eqref{eq:genjoint} can be rewritten as:
\begin{align}\label{eq:gen2}
P(y,s)= \sum_{\kappa,\kappa'} P(y|\kappa) P(s|\kappa') P(\kappa,\kappa').
\end{align}
 where  $P(\kappa,\kappa')= \sum_{\lambda} P(\kappa|\lambda)P(\kappa'|\lambda)P(\lambda)$   describes an arbitrary distribution over $\kappa$ and $\kappa'$, and 
where the distributions $P(y|\kappa)$ and $P(s|\kappa')$ range over the cyclic permutations of the distribution in Eq.~\eqref{eq:NCassignment}.

It follows that the distributions $P(y,s)$ with uniform marginal on $s$ that are achievable in a noncontextual model with the specified operational equivalences are all and only those lying in the intersection of the hyperplane defining the uniform marginal condition ($\forall s: \sum_y P(y,s) =1/5$) and the  hypercone consisting of the nonnegative hull of the twenty-five product distributions over $y$ and $s$,  $\{ P(y|\kappa) P(s|\kappa') \}_{\kappa,\kappa'}$,  one for each pair $\kappa,\kappa'$.  The product distributions in this set are the extremal rays of the hypercone.   Each 
 can be expressed as a $5\times 5$ matrix with rows indexed by $s$ and columns by $y$. Specifically, using the form of the extremal assignments, Eq.~\eqref{eq:NCassignment},  one infers that they are precisely those that can be obtained by cyclic permutations of the rows  and columns  of the matrix
\begin{align}\label{eq:vertex}
{M:=\frac{1}{5}}
\left(
\begin{array}{lllll}
1&  q^{-1} &  0 & 0 & q^{-1}  \\
q^{-1} & q^{-2}   & 0  & 0  & q^{-2}  \\
0 & 0  & 0 & 0 & 0 \\
0 & 0  & 0 & 0 & 0 \\
q^{-1} & q^{-2}   & 0  & 0  & q^{-2}
\end{array}
\right)\,.
\end{align}
We denote this twenty-five ray unbounded hypercone by $T_\text{unbounded}$. The intersection of $T_\text{unbounded}$ with the hyperplane defining the uniform marginal condition (i.e., the hyperplane where $\sum_y P(y,s) = 1/5$ for all $s$) defines a bounded polytope which we will denote by $T_\text{bounded}$.  $T_\text{bounded}$ is the polytope of noncontextually realizable joint distributions $P(y,s)$ with uniform marginal on $s$, for the given operational equivalences.  One could seek to identify the vertices and facet inequalities of $T_\text{bounded}$, but this is computationally nontrivial and it is not necessary for witnessing the failure of noncontextuality.  For any distribution $P(y,s)$ with uniform marginal on $s$, violation of any facet inequality of the hypercone $T_\text{unbounded}$ is sufficient to witness the fact that this distribution does not lie within $T_\text{bounded}$. Therefore, we here make use of some of the facet inequalities of $T_\text{unbounded}$ as our noncontextuality inequalities.

The polytope of noncontextually realizable {\em conditional} distributions $P(y|s)$, denoted $T^{\prime}_\text{bounded}$, is simply the image of $T_\text{bounded}$ under the map that rescales every joint distribution $P(y,s)$ with uniform marginal on $s$ by the factor 5, since for such distributions $P(y|s)=P(y,s)/P(s)= 5P(y,s)$. That is, $T^{\prime}_\text{bounded}$ is the intersection of $T_\text{unbounded}$ with the hyperplane where $\sum_y P(y|s) = 1$ for all $s$.   Since the hypercone $T_\text{unbounded}$ is invariant under nonnegative rescaling, we find that its facet inequalities constitute both noncontextuality inequalities for conditional distributions $P(y|s)$ or for joint distributions $P(y,s)$ with uniform marginals.
\footnote{It should be noted that we have conceptualized the prepare side of the prepare-measure scenario as a device wherein which of the five preparations is implemented is determined by the value of an {\em input} variable $s$.  In other words, $s$ has been conceptualized as a setting variable in a {\em multi-preparation}.  However, one can just as well conceptualize the prepare side of the prepare-measure scenario as consisting of a device that samples a value of $s$ at random, implements the associated preparation, and returns the value of $s$ as an {\em output}, so that $s$ is the outcome of a {\em source}.  The process by which such a source is obtained from the multi-preparation is simply an instance of flag-convexification.  The fact that one can evaluate noncontextual-realizability either in terms of the joint distribution $p(y,s)$ (for uniform marginal on $s$) rather than the conditional distribution $p(y|s)$ is an instance of the fact that one can always choose to flag-convexify the setting variable of the multi-preparation when assessing noncontextual realizability.  While different sources that define the same average state may be compatible or incompatible, the fact that one can always flag-convexify and thereby conceptualize the experiment as involving a {\em single} source implies that no incompatibility among sources is required for proofs of contextuality.  This point is considered in more detail in the companion article~\cite{companionincompatibility}.}

To identify facet inequalities of the hypercone $T_\text{unbounded}$ from its extremal rays, one can use linear programming algorithms.\footnote{We discover an inequality by minimizing the inner product of a variable vector with some ray outside of the hypercone subject to the restriction that vector have nonnegative inner product with each of the hypercone`s extremal rays. The inequality is seen to be facet-defining iff the vector space spanned with the subset of extremal rays which saturate the discovered inequality is exactly one dimension smaller than the vector space spanned by \emph{all} the hypercone's extremal rays.} Each facet inequality yields a noncontextuality inequality for this scenario.  
By applying standard such algorithms, we find that one such facet inequality is $\mathcal{C} \geq 0,$ where,  using the shorthand notation $p_{y|s}:=P(y|s)$, we can write $\mathcal{C}$ as 
\begin{align}
\mathcal{C}:= q \left(p_{1|0}+ p_{1|2}\right)+(q-1)p_{2|0}+p_{0|2}-(q+1)p_{1|1}\,. 
\end{align} Consequently, if a conditional distribution $P(y|s)$ violates the inequality $\mathcal{C}\geq 0$, the possibility of a noncontextual model for the given operational equivalences is ruled out.  One can easily verify that the extremal assignments (i.e., permutations of the rows and columns of Eq.~\eqref{eq:vertex}) satisfy this inequality. 

\begin{samepage}
Note that since the hypercone $T_\text{unbounded}$ is invariant under uniform rescaling, by substituting  $p_{y|s}\to 5  p_{y|s}:= P(y,s)$  we obtain the equally valid inequality $\mathcal{C}' \geq 0$, where
\begin{align}\label{NCinequality2}
\mathcal{C}':= q \left(p_{1,0}+ p_{1,2}\right)+(q-1)p_{2,0}+p_{0,2}-(q+1)p_{1,1}
\end{align}
which is satisfied by all $P(y,s)$ with uniform marginal $P(s)$ which admit a noncontextual model for the same operational equivalences.
\end{samepage}

One can generate forty-nine more facet inequalities of $T_\text{unbounded}$ by leveraging some of the symmetries of the problem. Since the set of extremal rays of $T_\text{unbounded}$ remains closed under the replacing $p_{y|s}$ with $p_{\pi(y)|\pi'(s)}$ where $\pi$ and $\pi'$ are cyclic permutations, or by replacing $p_{y|s}$  with $p_{s|y}$, it follows that any inequality generated by applying the inverse of such replacement to Eq.~(12) from the main text would also constitute a facet-defining inequality of $T_\text{unbounded}$.

For the prepare-measure scenario described in the main text and depicted in Fig.~(2) in the main text, the quantumly realized data table has elements $p_{y|s}= {\rm Tr}(\rho_s E_y)$.  
Computing these overlaps, the conditional distribution is described by the following $5\times 5$ matrix:
\begin{align}\label{eq:quantumpentagondata}
{M^{(q)}:= \frac{1}{10}}
\left(
\begin{array}{ccccc}
4&  1+q &  2-q & 2-q & 1+q  \\
1+q & 4   & 1+q  & 2-q  & 2-q  \\
2-q & 1+q  & 4 & 1+q & 2-q \\
2-q & 2-q  & 1+q & 4 & 1+q \\
1+q & 2-q   & 2-q  & 1+q  & 4
\end{array}
\right)\,.
\end{align}
One easily verifies that if one evaluates $\mathcal{C}$ in Eq.~(12) from the main text for $M^{(q)}$, one obtains the value $\frac{q^2-4}{10}\approx -0.138$  reported in Eq.~(14) of the main text, which is a violation of the noncontextuality inequality in Eq.~(13) from the main text.

 \end{document}